\documentclass[reprint,letterpaper,aps,prb,amsfonts,amsmath,twocolumn,superscriptaddress]{revtex4}
\usepackage{natbib}
\setcitestyle{super}
\usepackage[urlcolor=blue, hyperindex, colorlinks, bookmarks=true]{hyperref}
\usepackage{amsmath}
\usepackage{amsfonts}
\usepackage{amssymb}
\usepackage[english]{babel}
\usepackage{graphicx}  
\usepackage{dcolumn}   
\usepackage{bm}        
\usepackage{braket}

\usepackage{color}

\newcommand{\ground}{\vert \mathrm{g}\rangle}
\newcommand{\res}{\vert \mathrm{r}\rangle}
\newcommand{\bright}{\vert \mathrm{b}\rangle}
\newcommand{\dark}{\vert \mathrm{d}\rangle}
\newcommand{\done}{\vert \mathrm{d_1}\rangle}
\newcommand{\dtwo}{\vert \mathrm{d_2}\rangle}
\newcommand{\Qone}{\mathrm{Q}_1}
\newcommand{\Qtwo}{\mathrm{Q}_2}
\newcommand{\Qthree}{\mathrm{Q}_3}
\newcommand{\qone}{\vert \mathrm{q_1}\rangle}
\newcommand{\qtwo}{\vert \mathrm{q_2}\rangle}
\newcommand{\qthree}{\vert \mathrm{q_3}\rangle}
\newcommand{\A}{(\textbf{a})}
\newcommand{\B}{(\textbf{b})}
\newcommand{\C}{(\textbf{c})}
\newcommand{\D}{(\textbf{d})}

\newcommand{\avg}[1]{ \langle #1 \rangle }

\newcommand{\nocontentsline}[3]{}
\newcommand{\tocless}[2]{\bgroup\let\addcontentsline=\nocontentsline#1{#2}\egroup}

\begin{document}


\title{Studying Light-Harvesting Models with Superconducting Circuits}


\author{Anton Poto\v{c}nik}
\thanks{These authors contributed equally to this work}
\affiliation{Department of Physics, ETH Z\"urich, CH-8093, Z\"urich, Switzerland.}
\author{Arno Bargerbos}
\thanks{These authors contributed equally to this work}
\affiliation{Department of Physics, ETH Z\"urich, CH-8093, Z\"urich, Switzerland.}
\author{Florian~A.~Y.~N.~ Schr\"{o}der}
\affiliation{Cavendish Laboratory, University of Cambridge, J. J. Thomson Avenue, Cambridge CB3 0HE, United Kingdom.}
\author{Saeed A. Khan}
\affiliation{Department of Electrical Engineering, Princeton University, Princeton, New Jersey 08544, USA.}
\author{Michele C. Collodo}
\affiliation{Department of Physics, ETH Z\"urich, CH-8093, Z\"urich, Switzerland.}
\author{Simone Gasparinetti}
\affiliation{Department of Physics, ETH Z\"urich, CH-8093, Z\"urich, Switzerland.}
\author{Yves Salath\'e}
\affiliation{Department of Physics, ETH Z\"urich, CH-8093, Z\"urich, Switzerland.}
\author{Celestino Creatore}
\affiliation{Cavendish Laboratory, University of Cambridge, J. J. Thomson Avenue, Cambridge CB3 0HE, United Kingdom.}
\author{Christopher Eichler}
\affiliation{Department of Physics, ETH Z\"urich, CH-8093, Z\"urich, Switzerland.}
\author{Hakan E. T\"ureci}
\affiliation{Department of Electrical Engineering, Princeton University, Princeton, New Jersey 08544, USA.}
\author{Alex W. Chin}
\affiliation{Cavendish Laboratory, University of Cambridge, J. J. Thomson Avenue, Cambridge CB3 0HE, United Kingdom.}
\author{Andreas Wallraff}
\affiliation{Department of Physics, ETH Z\"urich, CH-8093, Z\"urich, Switzerland.}



\maketitle

\onecolumngrid
{\bf The process of photosynthesis, the main source of energy in the animate world, converts sunlight into chemical energy. The surprisingly high efficiency of this process is believed to be enabled by an intricate interplay between the quantum nature of molecular structures in photosynthetic complexes and their interaction with the environment. Investigating these effects in biological samples is challenging due to their complex and disordered structure. Here we experimentally demonstrate a new approach for studying photosynthetic models based on superconducting quantum circuits. In particular, we demonstrate the unprecedented versatility and control of our method in an engineered three-site model of a pigment protein complex with realistic parameters scaled down in energy by a factor of $10^5$. With this system we show that the excitation transport between quantum coherent sites disordered in energy can be enabled through the interaction with environmental noise. We also show that the efficiency of the process is maximized for structured noise resembling intramolecular phononic environments found in photosynthetic complexes.  }

\vspace{12pt}
\twocolumngrid

It is well accepted that the microscopic properties of all matter being composed of atoms and molecules are governed by the laws of quantum physics. At macroscopic scales, however, coherent quantum phenomena are frequently suppressed by the interaction with the environment. An intensely studied open question is, whether
quantum mechanics plays an important functional role in biological processes.
Examples of such processes are magneto reception in birds, olfaction and light harvesting, all studied in a field referred to as quantum biology \cite{Huelga2013,Lambert2013a,Scholes2017}. In particular, quantum coherent effects were observed in photosynthetic complexes by 2D electron spectroscopy at near ambient conditions \cite{Engel2007,Collini2010,Panitchayangkoon2010}, which stimulated both experimental and theoretical work on light harvesting.

In a photosynthetic process light is captured in a molecular complex acting as an antenna. The created excitation is then relayed towards a reaction center through a network of chlorophyll molecules forming  pigment protein complexes, such as the well studied Fenna-Matthews-Olson (FMO) complex \cite{Huelga2013,Lambert2013a}. At the reaction center the excitation enables the synthesis of energy-rich molecules, e.g.~adenosine triphosphate (ATP), relevant for supplying chemical energy throughout an organism. An excitation of an individual chlorophyll molecule is carried by a single chromophore whose highest occupied and lowest unoccupied molecular orbital can be approximated as a two-level system. The chlorophyll molecules form the sites of a network through which the individual excitations are transported.
The energy levels of the individual sites and the coupling between the sites are affected by both static and dynamic disorder, which in uniform systems suppresses energy transfer between sites.

High efficiency energy transport between disordered sites is suggested to be enabled through the interaction of the individual sites with vibrational modes of the protein scaffold into which the chlorophylls are embedded. A number of theoretical models have been developed to put this mechanism of efficient energy transport onto a solid footing \cite{Foerster1959,Olaya-Castro2008,Plenio2008,Rebentrost2009,Caruso2009}. In particular it has been suggested that interactions between the energy levels of the chlorophyll molecules and the highly-structured phononic environment of the protein scaffold enhance directed excitation transport \cite{Chin2012,Chin2013,Creatore2013,Rey2013}.

The direct verification of these models is challenging due to the intricate structure and the limited control obtainable over  photosynthetic complexes. Despite a number of theoretical studies, noise assisted energy transport (NAT) in biological systems has so far only been phenomenologically investigated on simple model systems with limited control over its parameters. Energy transport between two molecules placed on a substrate was studied with scanning tunneling microscopy  \cite{Imada2016}, using classical optics disorder was shown to break destructive interference and increase optical transmission \cite{Viciani2015,Viciani2016,Biggerstaff2016}, similarly, disorder in the coupling parameter was shown to increase energy flow using classical electronic circuits \cite{Leon-Montiel2015} and in genetically engineered molecular systems energy transport was controlled by adjusting inter-chromophoric distances \cite{Park2016,Hemmig2016}. Recently, a programmable nanophotonic processor was used to study transport properties in disordered systems \cite{Harris2017a}. 


\begin{figure*}[t]
\includegraphics[width=1\textwidth]{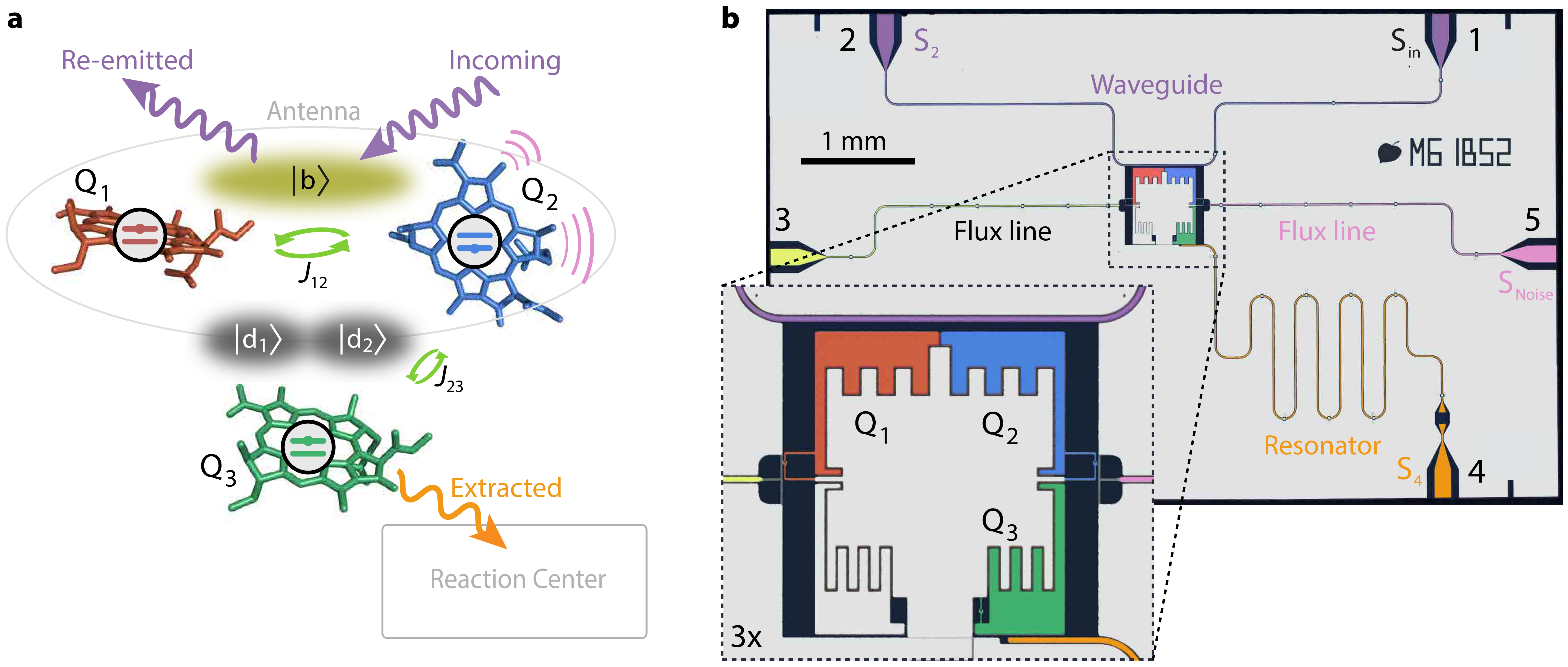}
\caption{{\bf Model system for light-harvesting and its  superconducting circuit realization.}   \A~Schematic of three coupled chlorophyll molecules hosting chromophores modeled as three qubits $\mathrm{Q}_{1,2,3}$ (red, blue, green). The strongly coupled $\Qone$ and $\Qtwo$ hybridize with coupling strength $J_{12}$ into a bright state $\bright$ (olive cloud) and dark states (dark gray clouds). $\Qtwo$ is coupled to $\Qthree$ with a coupling strength $J_{23}$ which form the dark hybridized states $\done$ and $\dtwo$. The incident, the re-emitted and the harvested radiation are indicated by arrows. Pink lines indicate that $\Qtwo$ is subjected to environmental noise.  \B~False-color micrograph of the superconducting circuit. Transmon qubits $\Qone$ and $\Qtwo$ (red, blue) are capacitively coupled to a transmission line (purple) and $\Qthree$ (green) is capacitively coupled to a high-emission-rate resonator (orange). Excitation radiation ($S_\mathrm{in}$) is applied through port 1 of the transmission line. The re-emitted ($S_2$) and extracted radiation ($S_4$) is detected at port 2 and 4, respectively. Low frequency noise with an engineered spectral density, modeling the environment ($S_\mathrm{noise}$), is applied to $\Qtwo$ via the flux line at port 5.}
\label{fig-1}
\end{figure*}

In this work we demonstrate the use of superconducting quantum circuits \cite{Schmidt2013,Houck2012} to test models describing important aspects of photosynthesis, such as photon absorption and noise assisted excitation transport, with unprecedented control in an engineered quantum system. 
We realize a small network of coherently coupled two-level systems with \textit{in-situ} tunable parameters interacting with an engineered environment, inspired by the proposal of Mostame {\it et al.} \cite{Mostame2012,Mostame2017}. Superconducting circuits are particularly well suited for this task, since versatile devices can be realized with a high degree of accuracy, and can be controlled and probed experimentally using well developed techniques \cite{daSilva2010}.

Here we experimentally investigate energy transport assisted by structured and unstructured environmental noise for coherent and incoherent excitation and demonstrate that its efficiency can approach unity. We also observe static coherences, even under incoherent excitation, and demonstrate good understanding of the full system dynamics.

{
\tocless
\section{Sample and Spectroscopic Characterization}
}

We implement a simplified model of a pigment protein complex consisting of three coupled chlorophyll molecules, labeled $\mathrm{Q}_{1,2,3}$ in Fig.~\ref{fig-1}a. The corresponding Hamiltonian is described in App.~\ref{sec:TheoryME}. This is the smallest system which incorporates all relevant elements, such as excitation trapping, energy mismatch, excitation delocalization, dark and bright states, necessary for studying noise assisted transport \cite{Plenio2008,Caruso2009,Huelga2013,Creatore2013,Rey2013,Chin2013}. Although current technology allows building larger systems capable of investigating for example the full FMO complex with 8 sites, we are convinced that it is a necessary first step to explore this novel approach on a simple model system. We realize two-level systems with individually tunable transition frequencies as transmon qubits \cite{Koch2007} in a superconducting circuit (Fig.~\ref{fig-1}b).
The dipole-dipole coupling between molecules $\Qone$ and $\Qtwo$ forms symmetric and antisymmetric, bright $\bright$ and dark $\dark$ state superpositions of the individual qubit excited states $\qone$ and $\qtwo$ \cite{May2004,Ferretti2016,VanLoo2013}. 
We realize the dipole-dipole interaction by direct capacitive coupling between qubits $\Qone$ and $\Qtwo$ (Fig.~\ref{fig-1}b). We excite the bright state $\bright$ through an open waveguide to which the two transmon qubits are coupled with equal strength \cite{VanLoo2013} modeling the excitation of the antenna part of the photosynthetic complex with photons propagating in free space. A third molecule $\Qthree$, coupled to $\Qtwo$, acts as a trap for the excitation which is subsequently extracted by transfer to the reaction center. In our circuit, this trapping site is realized as a third qubit which extracts the excitation from the system through its Purcell-like coupling to a transmission line resonator effectively acting as the reaction center. We model the interaction of molecule $\Qtwo$ with the environmental vibrational modes of the protein scaffold as fluctuations of its transition frequency. These fluctuations are induced by local magnetic fields acting on $\Qtwo$. While noise can be applied to all qubits, we chose to study local noise, such as the one induced by a local environment of a pigment protein. When applying noise to several sites we could study the effects  correlations in the noise have on the process.



\begin{figure}[t]
\begin{center}
\includegraphics[width=0.5\textwidth]{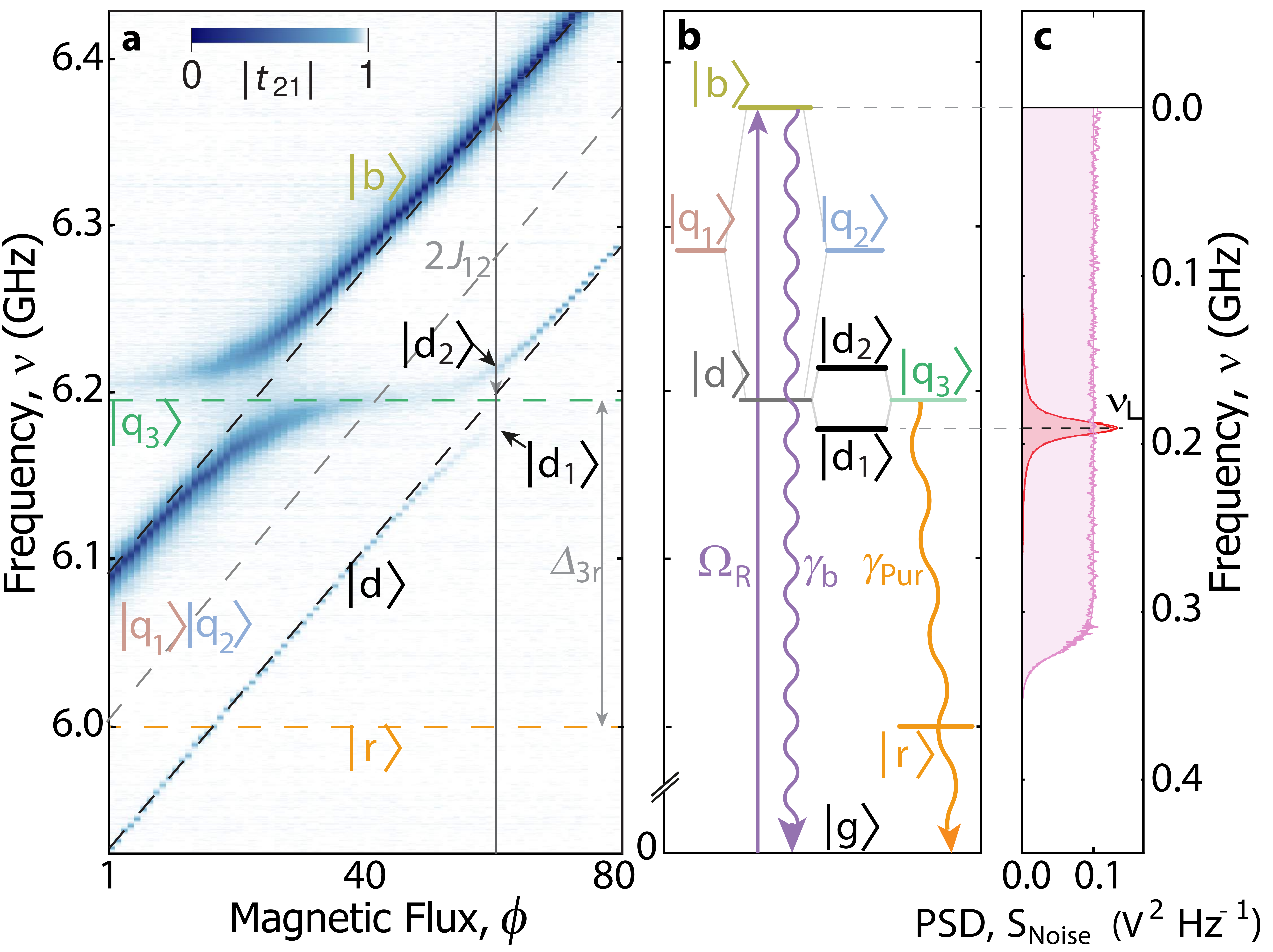}
\end{center}

\caption{{\bf Measured spectrum, energy levels and applied environmental noise spectra.} \A~Transmission spectra $\vert t_{21}(\omega)\vert $ of the three qubit system measured through the transmission line as a function of magnetic flux. Here and in the following, spectral features are labeled by the target state ($\qone,\qtwo,\qthree,\bright,\dark,\done,\dtwo, \res$) reached in spectroscopic experiments from the ground state $\ground$ of the system. \B~Energy level diagram at the magnetic flux indicated by a gray vertical line in \A. The resonant states $\qone$ and $\qtwo$ form bright $\bright$ and dark $\dark$ states. Furthermore, $\qthree$ is resonant with the $\dark$ state forming $\done$ and $\dtwo$ doublet. Solid downward arrows indicate decay channels, the upward arrow indicates excitation via the waveguide. \C~Measured power spectral density $S_\mathrm{Noise}$ of the environmental low-frequency noise applied to the flux line of $\Qtwo$. White noise with 325 MHz cutoff is depicted in pink and Lorentzian noise with central frequency $\nu_\mathrm{L}$ in red.}
\label{fig-2}
\end{figure}

We demonstrate a high degree of tunability of system parameters in a measurement of the frequency-dependent transmission coefficient $\vert t_{21}(\omega)\vert$ through the waveguide (see Fig.~\ref{fig-2}a) keeping the transition frequency of $\Qthree$ fixed at $\omega_3/2\pi = 6.198 \, \rm{GHz}$ and linearly sweeping the transition frequencies of $\Qone$ and $\Qtwo$ maintaining $\omega_1 = \omega_2$. We tune the qubit transition frequencies by magnetic fields applied using a coil and two flux lines shorted close to the SQUID loops of transmon qubits $\Qone$ and $\Qtwo$ (see App.~\ref{sec:SampleExperiment}). In this measurement, we observe that $\Qone$ and $\Qtwo$ form bright and dark states ($\bright, \dark$) with frequencies $\omega_\mathrm{b}$ and $\omega_\mathrm{d}$ separated by $2 J_{12}/2\pi = 173.4\,\rm{MHz}$ (see App.~\ref{sec:BrightDark}). The bright state linewidth $\gamma_\mathrm{b}/2\pi = 12.4$~MHz is consistent with the sum of the individual qubit radiative linewidths $\gamma_1/2\pi = 7.39$~MHz and $\gamma_2/2\pi = 6.57$~MHz dominated by the coupling to the waveguide. This indicates superradiance of the coupled two qubit system \cite{VanLoo2013,Mlynek2014b}. The subradiant dark state $\dark$ has a narrow linewidth of $\gamma_\mathrm{d}/2\pi = 0.29$~MHz limited by the residual asymmetry in $\Qone$ and $\Qtwo$ parameters (see App.~\ref{sec:SampleExperiment}) with a bright to dark state linewidth ratio of $\gamma_\mathrm{b}/\gamma_\mathrm{d} \approx 43$. Dark states have been suggested to improve the efficiency of biologically inspired photocells by protecting the excitation from re-emission \cite{Dong2012a,Creatore2013}. For $\omega_1/2\pi = \omega_2/2\pi = 6.285 \, \rm{GHz}$ (solid vertical line in Fig.~\ref{fig-2}a) the dark state $\dark$ coherently hybridizes with $\qthree$ forming a doublet $\done$ and $\dtwo$ split by $2J_\mathrm{d3}/2\pi = 37$~MHz consistent with the individual qubit couplings (see Fig.~\ref{fig-2}b and App.~\ref{sec:SampleExperiment}). We detune $\omega_3$ from the fixed frequency $\lambda/2$ resonator state $\res$  at $\omega_\mathrm{r}/2\pi = 6.00 \, \rm{GHz}$ by $\Delta_\mathrm{3r} = \omega_3 - \omega_\mathrm{r}$. This sets the radiative Purcell decay rate\cite{Sete2014} $\gamma_\mathrm{Pur}/2\pi \approx 20 \, \rm{MHz}$ of $\Qthree$ (see App.~\ref{sec:Purcell}) effectively modeling the energy extraction rate at the reaction center. As desired, all relevant microwave frequency system parameters are consistently scaled by a factor of $\sim 10^5$ relative to the optical frequency energy scales of the FMO complex\cite{Mostame2012}.

\vspace{32pt}
\tocless
\section{Excitation Transfer with Uniform White Noise}

\begin{figure}[t]
\includegraphics[trim={0cm 0cm 0cm 0cm},clip,width=0.5\textwidth]{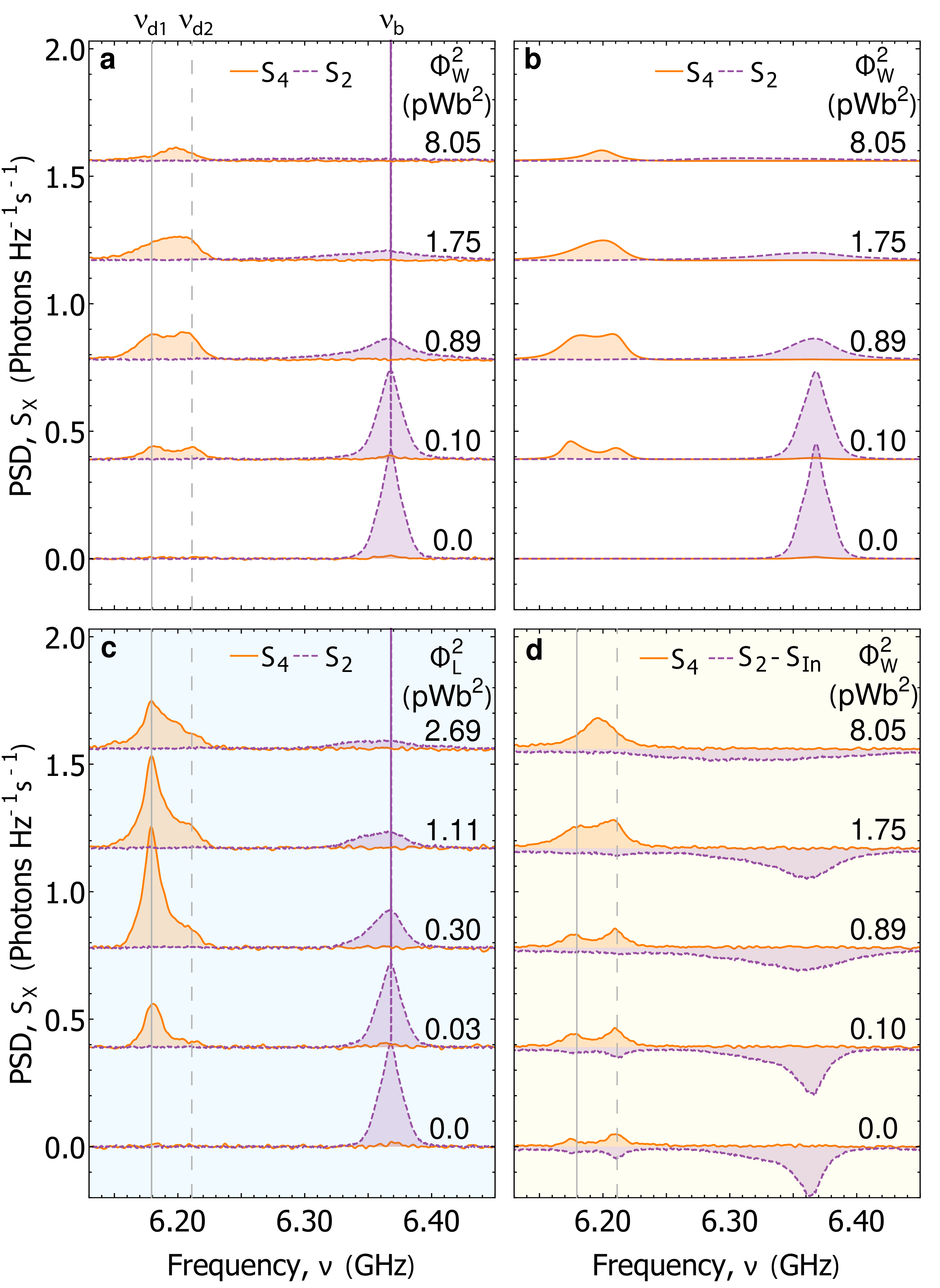}

\caption{{\bf Measured power spectral densities (PSD)} of radiation extracted from the resonator $S_\mathrm{4}(\omega)$ (solid orange lines) and re-emitted into the transmission line $S_\mathrm{2}(\omega)$ (dashed purple lines) for coherent excitation as a function of \A~white noise power $\Phi_\mathrm{W}^2$ and \C~Lorentzian noise power $\Phi_\mathrm{L}^2$. \B~Master equation calculations of $S_\mathrm{4}(\omega)$ and $S_\mathrm{2}(\omega)$ for coherent excitation as a function of $\Phi_\mathrm{W}^2$. \D~Measured PSD for incoherent excitation as a function of $\Phi_\mathrm{W}^2$. Dashed purple lines indicate the reduction of the measured PSD in incoherent microwave radiation $S_\mathrm{in}$, due to energy transfer through the system. PSD for different noise powers are displaced by 0.2 Photons s$^{-1}$Hz$^{-1}$. }

\label{fig-3}
\end{figure}

To study energy transfer through the circuit we tune $\Qone$ and $\Qtwo$ into resonance to form bright and dark states at frequencies $\omega_\mathrm{b}/2\pi = \nu_\mathrm{b} = 6.371\,\rm{GHz}$ and $\omega_\mathrm{d}/2\pi = 6.198 \, \rm{GHz}$. Qubit $\Qthree$ is tuned into resonance with the dark state $\omega_3=\omega_\mathrm{d}$ creating two resonances at $\omega_\mathrm{d1}/2\pi = \nu_\mathrm{d1} = 6.179 \,\rm{GHz}$ and $\omega_\mathrm{d2}/2\pi = \nu_\mathrm{d2} = 6.216 \, \rm{GHz}$. We coherently excite the bright state $\bright$ through port $1$ of the device with a continuous tone at frequency $\omega_\mathrm{b}$ and amplitude corresponding to a bright state Rabi frequency of $\Omega_\mathrm{R}/2\pi = 14$~MHz (see App.~\ref{sec:RabiCalibration}). We measure the power spectral density (PSD) $S_2(\omega)$ of the photons scattered along the waveguide into port $2$ of the device characterizing the re-emission from the absorption site. $S_\mathrm{2}(\omega)$ displays a narrow coherent peak at $\omega_\mathrm{b}$ due to elastically (Rayleigh) scattered photons and a broad resonance fluorescence spectrum with a width given by $\gamma_\mathrm{b}$  due to the inelastically scattered photons (bottom purple line in Fig.~\ref{fig-3}a). With increasing drive amplitude we observe a bright state Mollow triplet \cite{VanLoo2013}, see Fig.~\ref{fig:PowerCalibration}. Due to the energy mismatch of the bright state $\bright$ and the dark state doublet ($\done,\dtwo$) no excitations are transferred to qubit $\Qthree$ and thus no photons are detected at the resonator port $4$, as shown by the vanishing PSD $S_4(\omega)$ at $\Phi_\mathrm{W}^2 = 0$~pWb$^2$ (bottom orange line in Fig.~\ref{fig-3}a). In our model system, no energy is transferred from the antenna to the reaction center in the absence of environmental noise.

To engineer a broad environmental noise spectrum, such as the one generated by the combination of background thermal noise and overlapping vibrational modes of the protein scaffold present in light harvesting systems \cite{Mostame2012}, we apply white Gaussian noise to port $5$ inducing frequency fluctuations in $\Qtwo$. The broad Markovian noise has a power spectral density of adjustable amplitude, constant up to a cutoff frequency of $325 \, \rm{MHz}$, characterized by its integrated flux noise power $\Phi_\mathrm{W}^2$ at qubit $\Qtwo$ (see App.~\ref{sec:Noise} and Fig.~\ref{fig-2}c). We note that applying synthesized noise to $\qtwo$ effectively creates a classical environment that can be described by the Haken-Strobl-Reineker model \cite{Haken1972,Haken1973} for white noise. Applying classical, as opposed to quantum noise, offers a unique possibility to engineer noise with controllable power spectral density capable of creating environments that approximate those of pigment-protein complexes \cite{Mostame2012} without increasing complexity of the device design.

Even for small applied noise power we observe energy transfer from the bright $\bright$ to the dark state doublet $\done,\dtwo$ indicated by two resonances at frequencies $\omega_\mathrm{d1}$ and $\omega_\mathrm{d2}$ in the detected PSD $S_4(\omega)$ (orange line at $\Phi_\mathrm{W}^2 = 0.1 \, \rm{pWb^2}$ in Fig.~\ref{fig-3}a). The excitation transport is enabled by those frequency components of the noise spectrum that bridge the energy difference $2J_{12}\pm J_\mathrm{d3}$ between bright $\bright$ and dark states $\done,\dtwo$. We have verified this aspect by reducing the bandwidth of the noise to below that energy difference in which case no energy transfer is observed. The emission linewidths, i.e.~the emission rates of $\done$ and $\dtwo$ into the resonator are determined by the Purcell decay rate $\gamma_\mathrm{Pur}$ (see App.~\ref{sec:Purcell}). The well resolved doublet in the detected spectrum $S_4(\omega)$ indicates that static coherences of the underlying quantum network are observable in noise induced transport. Based on the observations of the doublet, we expect beatings with frequency $2 J_\mathrm{d3}$ to be observable in temporally resolved measurements of the power at the extraction site.

With increasing applied noise power $\Phi_\mathrm{W}^2$, the power spectrum $S_2(\omega)$ of the resonance fluorescence of the bright state $\bright$ broadens due to the pure dephasing induced by the noise. From this measurement we determine the bright state pure dephasing rate $\gamma_\phi^\mathrm{b}$ in dependence on the applied white noise power $\Phi_\mathrm{W}^2$ (see App.~\ref{sec:DephasingCalibration}).
The extracted power $S_4(\omega)$ first increases with increasing noise power $\Phi_\mathrm{W}^2$ while the doublet remains resolved. At noise powers above $\Phi_\mathrm{W}^2 \approx 2$~pWb$^2$ the observed doublet transforms into a single resonance marking a crossover from the strong-coupling regime ($2J_\mathrm{d3}\gtrsim\gamma^\mathrm{b}_\phi$) to the weak-coupling regime ($2J_\mathrm{d3}\lesssim\gamma^\mathrm{b}_\phi$) where the remaining resonance stems from the incoherently excited $\qthree$ state (see App.~\ref{sec:WhiteNoise}). Beyond this threshold, the extracted power decreases. For this simple situation of only three sites and Markovian noise, all essential features of the experimentally observed power spectra are consistent with both Lindblad master equation and Bloch-Redfield calculations (see Fig.~\ref{fig-3}b and App.~\ref{sec:Theory}).

Integrating the measured power spectral densities $S_\mathrm{4}(\omega)$ and $S_\mathrm{2}(\omega)$ while omitting contributions from elastic (Rayleigh) scattering, we find that the total power re-emitted from the bright state into the waveguide in forward direction $P_\mathrm{2}$ decreases monotonically as a function of applied noise power $\Phi_\mathrm{W}^2$ (open purple squares in Fig.~\ref{fig-4}a). In contrast, the total power detected at the extraction site $P_\mathrm{4}$ first increases rapidly with $\Phi_\mathrm{W}^2$, exhibits a pronounced maximum and then decreases again (open orange diamonds in Fig.~\ref{fig-4}a). The increase for small dephasing rates is a consequence of noise-induced incoherent transitions between bright and dark states \cite{Creatore2013,Fruchtman2016} allowing the system to overcome the energy mismatch.

From the integrated powers we calculate the transfer efficiency of the excitation from the absorption site to the extraction site as $\eta = {P_\mathrm{4}}/({P_\mathrm{4} + 2P_\mathrm{2}})$. The factor 2 accounts for the bidirectional character of the bright state resonance fluorescence \cite{VanLoo2013}, i.e.~equal powers are emitted in forward and backward direction, while we detect only in forward direction. The transport efficiency $\eta$ (green circles in Fig.~\ref{fig-4}a) shows a rapid increase from zero, a broad maximum of $\eta_\mathrm{W}^\mathrm{max} = 39\%$ and then a slow decrease with increasing pure dephasing rate $\gamma^\mathrm{b}_\phi$ (see top axis in Fig.~\ref{fig-4}a), which are the characteristic features of noise assisted transport \cite{Olaya-Castro2008,Plenio2008,Rebentrost2009}. The decrease in efficiency above an optimal noise power is due to dephasing-induced population localization, also referred to as the quantum Zeno effect \cite{Olaya-Castro2008,Plenio2008,Rebentrost2009}.

The measured integrated powers and hence the efficiency are in good agreement with results from master equation simulations (solid lines in Fig.~\ref{fig-4}a). Using rate equations (see App.~\ref{sec:TheoryRE}) we show that the maximal efficiency is $\eta^\mathrm{max}_\mathrm{W} \approx (1 - \gamma_\mathrm{b}/\gamma_\mathrm{Pur})$  approaching 100\% for $\gamma_\mathrm{b}\ll\gamma_\mathrm{Pur}$. Although small $\gamma_\mathrm{b}$ maximizes the transfer efficiency \cite{Rebentrost2009}, the total extracted power at optimal applied noise is proportional to $\gamma_\mathrm{b}$. Therefore for practical light harvesting applications one may choose to maximize output power while compromising on efficiency \cite{Creatore2013,Fruchtman2016}.
Similarly, we have not chosen a smaller $\gamma_\mathrm{b}$ in our experiment to maximize the efficiency, but opted for a larger extracted power to achieve a high signal-to-noise ratio at an acceptable integration time.
Finally, we note that in our data the maximum efficiency $\eta^\mathrm{max}_\mathrm{W}$ occurs at the strong-to-weak coupling crossover $2J_\mathrm{d3}\approx\gamma_\phi^\mathrm{b}$. At this point, the transfer rate $\gamma_\phi^\mathrm{b}$ between $\bright$ and $\dark$ is comparable to the transfer rate between $\dark$ and $\qthree$,  which is approximately given by $2J_\mathrm{d3}^2/\gamma_\phi^\mathrm{b}$ (see App.~\ref{sec:TheoryRE}). This experimentally demonstrates the interplay between quantum coherent effects and classical dephasing enhancing excitation transport.

\vspace{32pt}
\tocless
\section{Excitation Transfer with Lorentzian Environment}

It has been conjectured that structured environmental noise, such as the one originating from long lived vibrational modes of chlorophyll molecules in photosynthetic complexes, can further enhance the energy transfer efficiency between the disordered molecular sites of the network in a scenario known as the phonon antenna mechanism \cite{Chin2012,Huelga2013}. To demonstrate this concept, we apply environmental noise with Lorentzian PSD, characterized by its central frequency $\nu_\mathrm{L}$, its width $\Delta\nu_\mathrm{L}$ and amplitude (Fig.~\ref{fig-2}c), to qubit $\Qtwo$. We select a fixed bandwidth $\Delta\nu_\mathrm{L} = 10$~MHz which in good approximation corresponds to the scaled linewidth of the environmental noise expected from vibrational modes in natural photosynthetic complexes \cite{Mostame2012,Rey2013}. Since $\Delta\nu_\mathrm{L}$ is comparable to the decay rates $\gamma_\mathrm{b}$ and $\gamma_{\rm{Pur}}$ the spectral properties of the noise effectively create a non-Markovian environment. Initially, we choose the Lorentzian central frequency $\nu_\mathrm{L} = 190$~MHz to be resonant with the $\bright$ to $\done$ frequency difference $\Delta_\mathrm{b,d1}$. The qualitative features of the measured power spectral densities $S_2(\omega)$ and $S_4(\omega)$ and their dependence on the integrated applied noise power $\Phi_\mathrm{L}^2$ (Fig.~\ref{fig-3}c) are comparable  to the white noise case (Fig.~\ref{fig-3}a) with some distinct differences. As a direct consequence of applying Lorentzian noise resonant at $\Delta_\mathrm{b,d1}$, $S_\mathrm{4}(\omega)$ exhibits a strong resonance exactly at $\omega_\mathrm{d1}$ and a weaker resonance at $\omega_\mathrm{d2}$. The excitation transfer can be interpreted as a two-photon process \cite{Li2012e} absorbing one photon from the coherent excitation field at frequency $\omega_\mathrm{b}$ and emitting one photon into the environmental noise field at frequency $\Delta_\mathrm{b,d1}$ or $\Delta_\mathrm{b,d2}$. This effectively creates a transition from the joint ground state through the bright state $\bright$ into the dark states $\done$ or $\dtwo$ from which energy is extracted.

When we sweep the Lorentzian noise center frequency over a broad range from $\nu_\mathrm{L} = 0$ to $300 \, \rm{MHz}$ at weak noise power $\Phi_\mathrm{L}^2 = 0.016$~pWb$^2$, we observe two well resolved maxima in transferred power $P_4$ when the noise is resonant with the bright to dark state frequency differences $\Delta_\mathrm{b,d1}$ and $\Delta_\mathrm{b,d2}$ (Fig.~\ref{fig:LorCentFreqSweep}). 
In contrast, no energy transfer is observed when the noise is far detuned. Both observations clearly demonstrate the strong sensitivity of the energy transfer on the spectral properties of the environmental noise and indicate that the noise assisted transport is enabled by a narrow part of the noise PSD that matches the energy gap otherwise blocking the energy transfer in the system. This observation is consistent with stochastically averaged master equation simulations (see Fig.~\ref{fig:StochasticSim} and App.~\ref{sec:StochasticME}). 
We note that despite applying classical noise the dynamics induced by Lorentzian environment cannot be simulated with the HSR approach or even its extension for colored noise \cite{Warns1998, Barvik2000}, due to its strong non-Markovian  character. Therefore, a full quantum numerical simulation is required at the cost of significant computational effort.

\begin{figure*}[t]
\includegraphics[trim={0cm 0cm 0cm 0cm},clip,width=1.0\textwidth]{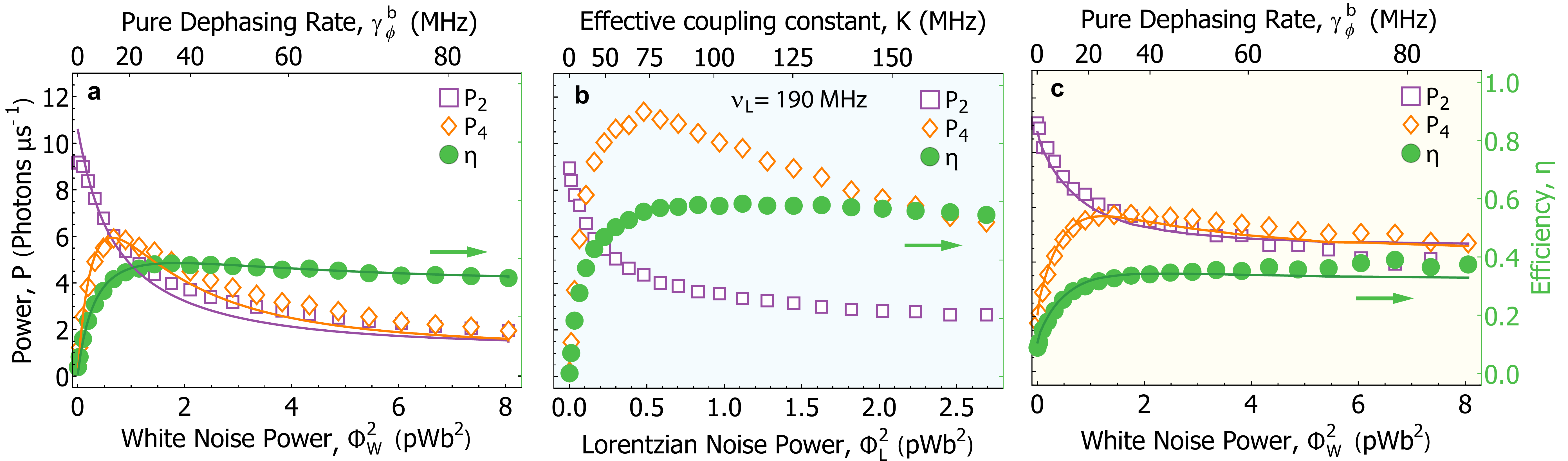}

\caption{{\bf Total extracted power $P_\mathrm{4}$ (orange diamonds), re-emitted power $P_\mathrm{2}$ (purple squares) and transport efficiency $\eta$ (green circle)} for {\A } coherent excitation and white noise environments, {\B } coherent excitation and Lorentzian noise environments and {\C } incoherent excitation and white noise environments. Data is plotted as a function of white noise power $\Phi_\mathrm{W}^2$ or equivalently bright state dephasing rate $\gamma^\mathrm{b}_\phi$ for {\A } and {\C } and as a function of Lorentzian noise power $\Phi_\mathrm{L}^2$ or equivalently effective qubit-environment coupling constant $K$ for \B. Solid lines are results of master equation simulations (see App.~\ref{sec:TheorySim}).}

\label{fig-4}
\end{figure*}

For Lorentzian environmental noise, the integrated detected powers also display the characteristic properties of noise assisted transport as a function of $\Phi_\mathrm{L}^2$ (Fig.~\ref{fig-4}b) as discussed before for white noise (Fig.~\ref{fig-4}a). However, we note that the extracted power $P_\mathrm{4}$ is almost twice as large at the same bright state excitation amplitude (Fig.~\ref{fig-4}a,b) leading to increased maximal efficiency $\eta_\mathrm{L}^\mathrm{max} = 58$\%. This indicates that structured environmental noise matching internal energy differences of the quantum network
indeed enhances the efficiency of the energy transfer. Estimating the effective qubit-environment coupling constant $K$ (see Fig.~\ref{fig-4}b and App.~\ref{sec:EffectiveQubitPhononCoupling}) we observe that the maximum in efficiency coincides with $K/2\pi \approx 100$~MHz which is comparable to the inter-qubit coupling constant $J_{12}$. This demonstrates that strong coupling between qubit and phononic modes is required to achieve maximal energy transport. Such a situation has been suggested to enhance the transport in cyanobacterial light-harvesting proteins, allophycocyanin and C-phycocyanin \cite{Womick2011}.


We have also applied a coherent tone with controlled frequency $\nu_\mathrm{c}$ and amplitude to qubit $\Qtwo$ through the environmental channel (port 5) (see App.~\ref{sec:CoherentMod}) and observe even larger extracted powers at the same bright state $\bright$ input field amplitude and near unit transfer efficiency (see Fig.~\ref{fig:CoherentFLInt}c). While this case does not occur in natural light harvesting since the environment cannot be fully coherent, it represents an interesting limiting case. In natural light harvesting systems exposed to incoherent environments, the highest transfer efficiency is realized for structured environmental noise with a narrow spectral density peaked at the frequency of the internal energy mismatch between the sites of the network. Any excess spectral width of the environmental noise leads to additional dephasing (see App.~\ref{sec:TheoryLorNoise}), which in turn reduces the absorption and energy transfer efficiency. These aspects are clearly demonstrated by the presented set of experiments comparing energy transfer with white and Lorentzian noise spectral density.

\vspace{42pt}
\tocless
\section{Incoherent Excitation}\

In the final set of experiments we excite the qubit system with incoherent microwave radiation to mimic excitation of biological pigment protein complexes with sunlight. We engineer $0.95$~GHz broad incoherent microwave radiation centered at $\omega_\mathrm{B}$ that spans over all qubit transition frequencies (see App.~\ref{sec:IncoherentExcitationA}). The incoherent microwave power integrated over the bright state spectrum was adjusted to be equal to the drive power used for the case of coherent excitation. In this experiment we study the transport of incoherently created excitations as a function of applied white noise power. 

Since it is not possible to distinguish between the broad incoming incoherent and the re-emitted radiation at port 2 we plot in Fig.~\ref{fig-3}d the difference between detected power spectrum $S_2$ at port 2 and the separately measured incoherent radiation spectrum $S_\mathrm{In}$. The difference spectrum corresponds to the sum of the absorbed and the re-emitted spectrum $\tilde{S}_2$.

When increasing the applied white noise power we observe that $S_4(\omega)$ and [$S_2(\omega)-S_\mathrm{In}(\omega)$] show general features (Fig.~\ref{fig-3}d) similar to the ones observed for coherent excitation. However, in case of incoherent excitation a finite power ($P_4 = 0.9$~Photons $\mu \mathrm{s}^{-1}$) is extracted at port 4 even in the absence of applied environmental white noise ($\Phi_\mathrm{W}^2 = 0$). The observed extracted power is a result of direct excitation of the dark state, due to its finite coupling to the open waveguide. When the dark state is not completely dark simultaneous incoherent excitation of bright and dark states reduces coherence between $\Qone$ and $\Qtwo$ and therefore increases dephasing of the system. Existence of dark states in photosynthetic complexes can therefore help protect the system against dephasing induced by incoherent excitation.
The observation of the $\done$, $\dtwo$ doublet in $S_2(\omega)$ (Fig.~\ref{fig-3}d) demonstrates that static coherences can be observed for incoherent excitation, i.e. even in the absence of coherent sources, as long as the coherent coupling between the sites is larger than the total dephasing.

The maximum of the measured efficiency $\eta_\mathrm{W,inc.}^\mathrm{max} = 38\%$ is smaller but comparable to the case of coherent excitation with the maximum shifted towards higher applied environmental white noise powers. Comparable efficiencies are consistent with rate equation descriptions where the efficiency is independent of the spectral and coherence properties of the excitation. On the other hand, the extracted power $P_4$ is by more than a factor 2 larger at the highest environmental noise power ($\Phi_\mathrm{W}^2 = 8.05$~pWb$^2$) compared to the coherent excitation case. This indicates that absorption of the incoherent photons is not as strongly affected by environmental dephasing as for coherent excitation, due to a persistent overlap between broadened bright state spectrum and spectrum of the incoherent irradiation. Similar conclusion can be made for energy transport induced by Lorentzian noise for incoherent excitation (see App.~\ref{sec:IncoherentExcitationLorentzian}).

\vspace{32pt}
\tocless
\section{Conclusions}



In a proof of concept experiment we studied models of photosynthetic processes using superconducting quantum circuits. With a system of three coupled qubits we demonstrated how the interplay of quantum coherence and environmental interactions affects energy transport in a system with excellent control achievable over all relevant parameters. We expect this approach to be extensible to study other relevant aspects of light harvesting, such as time-resolved dynamics of the coherent excitation transfer; the role of quantum environments realizable in electronic circuit models as low frequency quantum harmonic oscillators \cite{Mostame2012,Creatore2013}; and scaling to systems with a larger number of coherent sites such as the FMO complex. Furthermore, we expect similar approaches to be applicable not only to study light-harvesting processes but also other interesting aspects of quantum biology such as the sense of smell in animals and humans and magneto reception in birds \cite{Huelga2013,Lambert2013a}. It could also be interesting to evaluate the potential of the techniques presented here to model processes in quantum chemistry and search for potential future applications of related methods to support, for example, the design of catalysts, e.g.~for nitrogen fixation, or bio-molecular compounds for drug development.

\vspace{12pt}
{\bf \noindent Acknowledgements}
We are grateful to G. Blatter and D. Vion for helpful feedback on the manuscript and T. Walter and P. Kurpiers for valuable discussions. Work of H.E.T. and S.K. was supported by the US Department of Energy, Office of Basic Energy Sciences, Division of Materials Sciences and Engineering, under Award No. DE-SC0016011. A.W.C. and F.A.Y.N.S. acknowledge support from the Winton Programme for the Physics of Sustainability. F.A.Y.N.S. also acknowledges support by the Engineering and Physical Sciences Research Council (EPSRC). Work of A.P., A.B., M.C., S.G., Y.S., C.E. and A.W. was supported by ETH Z\"urich.

\vspace{12pt}
{\bf \noindent Competing Interests}
The authors declare that they have no
competing financial interests.

\vspace{12pt}
{\bf \noindent Correspondence}
Correspondence and requests for materials
should be addressed to A.P.\\~(email: anton.potocnik@phys.ethz.ch).

\vspace{12pt}
{\bf \noindent Author Contributions}
AP, AB and MCC designed the sample, performed the experiment and analyzed the data. SG fabricated the sample. The FPGA firmware was implemented by YS. FAYNS, CC and AWC performed numerical simulations with the uniform environment and contributed to the experimental set-up. SAK and HET performed numerical simulations with the structured environment. AP and AW co-wrote the manuscript. CE commented on the manuscript.  All authors contributed to the manuscript preparation. The project was led by AP and AW.


%

\begingroup
\def\refname{}
\def\bibname{}

\endgroup



\newpage

\appendix

\onecolumngrid
\begin{center}
{\bf \huge Appendix}
\end{center}

\vspace{24pt}
\twocolumngrid

\tableofcontents

\addtocontents{toc}{\setcounter{tocdepth}{0}}

\setcounter{figure}{0}
\renewcommand{\thefigure}{S\arabic{figure}}

\section{Sample and Experimental Setup}
\label{sec:SampleExperiment}
%

The three qubits are implemented with a grounded design, similar to X-mon qubits \cite{Barends2013}, to minimize the unwanted capacitive coupling between $\Qone$ and $\Qthree$. Their arrangement (see Fig.~\ref{fig-1}b in the main text) yields capacitive coupling rates of $J_{12}/2\pi = 83.6$~MHz, $J_{23}/2\pi = 33.4$~MHz and an order of magnitude smaller $J_{13}/2\pi = 3.67$~MHz. All reported qubit parameters are determined at the qubit transition frequency of $\omega/2\pi = 6.28$~GHz. When extracting individual qubit parameters the other two qubits are detuned by at least 1.5 GHz. The measured maximum transition frequencies between the ground $\vert g\rangle$ and first excited state $\vert e\rangle$ are $\omega_\mathrm{1}^\mathrm{max}/2\pi = 6.948$~GHz, $\omega_\mathrm{2}^\mathrm{max}/2\pi = 6.694$~GHz and $\omega_\mathrm{3}^\mathrm{max}/2\pi = 7.271$~GHz for the three qubits and their anharmonicities of the first-to-second excited state are $\alpha_1/2\pi = -140$~MHz, $\alpha_2/2\pi = -142$~MHz and $\alpha_3/2\pi = -137$~MHz. The spectroscopically measured pure dephasing rates of $\Qone$ and $\Qtwo$ are $\gamma_\phi^{(1)}/2\pi = 115$~kHz and $\gamma_\phi^{(2)}/2\pi = 82$~kHz. $\Qone$ and $\Qtwo$ are coupled to an open waveguide (transmission line) with coupling rates $\gamma_1/2\pi = 6.57$~MHz and $\gamma_2/2\pi = 7.39$~MHz. $\Qthree$ is coupled to a $\lambda/2$ resonator with coupling coefficient $g/2\pi \approx 90$~MHz. The uncoupled resonator has a fundamental frequency of $\omega_\mathrm{r}/2\pi = 6.00$~GHz and a loaded quality factor of $Q_\mathrm{L}  \approx Q_\mathrm{ext} \approx 55$ dominated by the external coupling.
The transition frequencies of the three qubits are tuned by magnetic flux $\Phi_i$,\cite{Koch2007}
\begin{eqnarray}
\omega_i(\Phi_i) \simeq (\omega_i^\mathrm{max}-\alpha_i)\sqrt{\left\vert \cos(\pi\Phi_i/\Phi_0)\right\vert}+\alpha_i,\label{eq:QubitTune}
\end{eqnarray}
where $\Phi_0$ is the flux quantum, for $i = 1-3$ qubits. Magnetic flux is generated by applying DC currents to two flux lines (FL1, FL2) located close to the SQUID loops of $\Qone$ and $\Qtwo$ and to a superconducting coil coupled globally to all three qubits. Individual qubit frequency control is obtained by inverting the flux coupling matrix and applying appropriate currents.

\begin{figure*}
\includegraphics[trim={0cm 0cm 0cm 0cm},clip,width=1\textwidth]{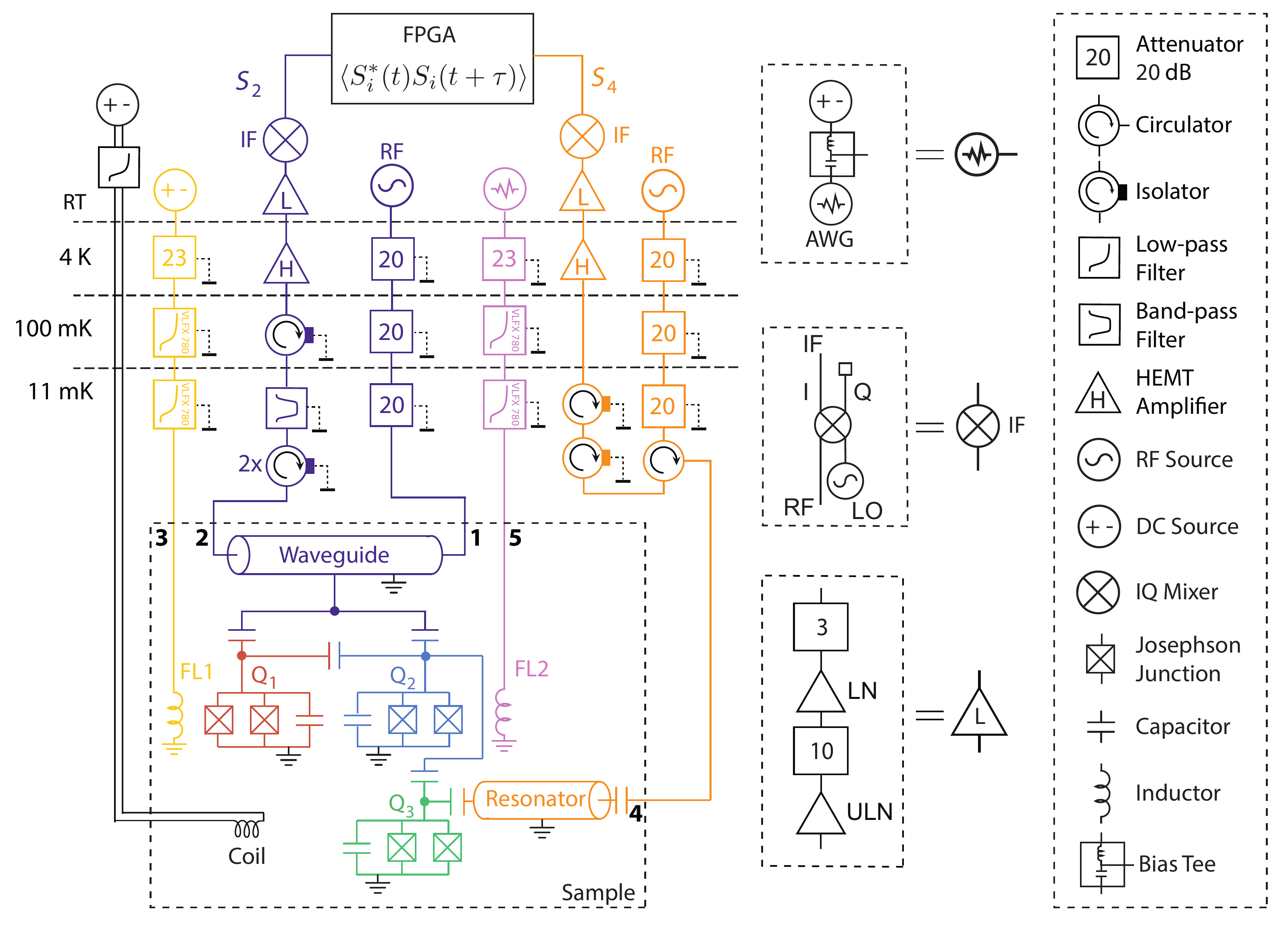} 
\caption{{\bf Schematic of the experimental setup}
 with complete wiring of electronic components inside and outside of the dilution refrigerator with the same color code as used for the sample in Fig.~\ref{fig-1}b of the main text.}
\label{fig:ExpSetupDiagram}
\end{figure*}

Coherent microwave radiation (RF) generated at room temperature by a commercial source is thermalized and attenuated at the 4~K, 100~mK and 11~mK stages of dilution refrigerator and applied to the sample at port 1 of the waveguide (see schematic diagram in Fig.~\ref{fig:ExpSetupDiagram}). Radiation emitted from the waveguide and from the resonator is amplified with high-electron-mobility transistor (HEMT) amplifiers at 4~K followed by a chain of ultralow-noise (ULN) and low-noise (LN) amplifiers at room temperature. Three isolators are inserted between the sample and the HEMT amplifier to suppress the amplifier input noise propagating back to the sample. The radiation emitted from port 2 of the waveguide is filtered with a band-pass filter (BPF). The amplified signals are down-converted to an intermediate frequency (IF) of 250 MHz with an IQ mixer using a local oscillator (LO) tone and then digitized with an analog-to-digital converter (ADC). The digital signal in the measurement bandwidth of 250 MHz is then processed by a field-programmable-gate-array (FPGA) which determines the amplitude and the power spectral density of the signal \cite{daSilva2010}. Typically $2^{24} \approx 16\cdot10^6$ samples are collected in about 15~min to obtain a single power spectral density $S(\omega)$ measurement. The low frequency noise is generated by an arbitrary waveform generator (AWG) and combined with the DC bias using a bias tee with a low-frequency cutoff of 5~kHz and then applied to FL2 (see Fig.~\ref{fig:ExpSetupDiagram}). Instead of an AWG we could use, for example, a heated resistor to generate low frequency white noise. However, AWG offers a unique advantage of {\em in-situ} control of both amplitude and shape of the noise power spectrum and has been used previously to create quasi thermal noise in circuit QED experiments \cite{Fink2010}.


\section{Description of Symmetric and Antisymmetric superposition of $\Qone$ and $\Qtwo$ }
\label{sec:BrightDark}

\begin{figure}[t]
\begin{center}
\includegraphics[trim={0cm 0cm 0cm 0cm},clip,width=0.49\textwidth]{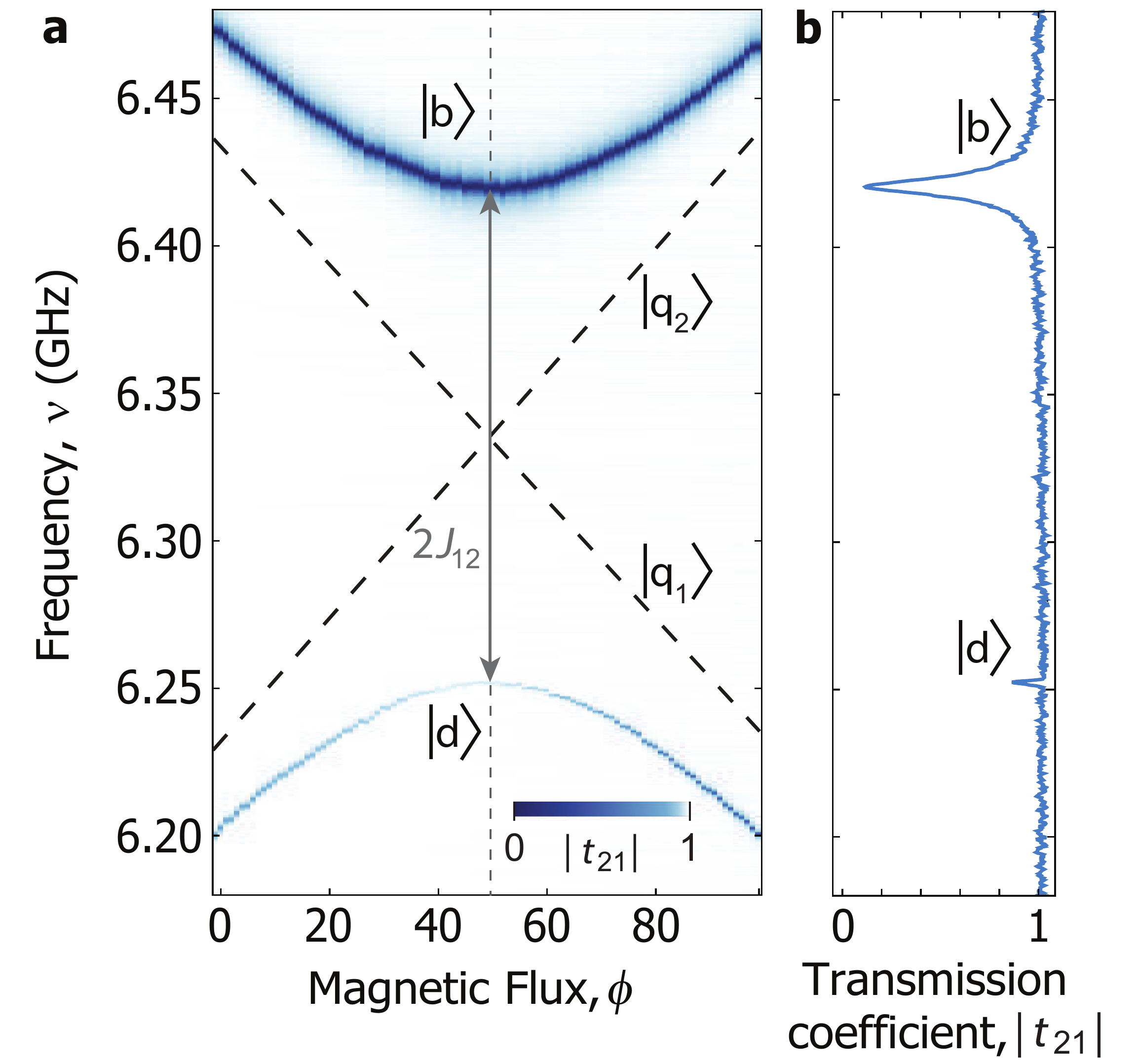} 
\end{center}
\caption{{\bf $\Qone$, $\Qtwo$ avoided crossing.}
\A~Measured frequency dependent transmission coefficient $|t_{21}|$ of the waveguide as a function of magnetic flux. $\qone$ and $\qtwo$ transition frequencies linearly cross (black dashed lines). $\qthree$ is detuned by more than 1.5~GHz. \B~$|t_{21}|$ measurement as a function of frequency $\nu$ at $\phi = 49$ [vertical dashed line in \A] where $\qone$ and $\qtwo$ are maximally hybridized.}
\label{fig:DarkBrightAnti}
\end{figure}

On resonance, $\Qone$ and $\Qtwo$ excited states $\qone$ and $\qtwo$ form symmetric and antisymmetric (bright and dark) states $\bright = \left(\qone + \qtwo\right)/\sqrt{2}$ and $\dark = \left(\qone - \qtwo\right)/\sqrt{2}$. The formation of the hybridized states is observed as an avoided crossing in measurements of the transmission coefficient $|t_{21}|$ through the waveguide for magnetic flux chosen such that the bare frequencies of $\Qone$ and $\Qtwo$ cross, indicated by the black dashed lines in Fig.~\ref{fig:DarkBrightAnti}a. Fully hybridized $\bright$ and $\dark$ states at zero detuning between $\Qone$ and $\Qtwo$ ($\Delta_{12} = 0$) are separated in energy by $2J_{12}$ (Fig.~\ref{fig:DarkBrightAnti}b). We fit the measured transmission coefficient at the point of maximal hybridization using the expression \cite{Astafiev2010}
\begin{equation}
t = 1 - \frac{\gamma_r}{\gamma_r+2\gamma_\phi}\frac{1-\frac{i\Delta}{\gamma_r/2+\gamma_\phi}}{1+\left(\frac{\Delta}{\gamma_r/2+\gamma_\phi}\right)^2+\frac{\Omega_\mathrm{R}^2}{\gamma_r(\gamma_r/2+\gamma_\phi)}},
\label{eq:lineshapeasta}
\end{equation}
with detuning $\Delta = \omega_i - \omega_\mathrm{in}$, excitation frequency $\omega_\mathrm{in}$ and Rabi rate $\Omega_\mathrm{R}$. From the fit we extract the radiative decay rate $\gamma_r$, dominated by the coupling rate to the waveguide, and the pure dephasing rate $\gamma_\phi$ for both states. Here we neglect the non-radiative decay. We find a higher frequency bright state coupling rate of $\gamma_\mathrm{b}/2\pi = 12.44$ MHz and pure dephasing rate of $\gamma_\phi^\mathrm{b}/2\pi = 0.38$ MHz and a lower frequency subradiant state coupling rate of $\gamma_\mathrm{d}/2\pi = 0.29$~MHz and comparable pure dephasing rate of $\gamma_\phi^\mathrm{d}/2\pi = 0.55$~MHz.

\section{Characterization of $\Qthree$ and its Purcell Decay}
\label{sec:Purcell}

\begin{figure}[t]
\begin{center}
\includegraphics[trim={0cm 0cm 0cm 0cm},clip,width=0.47\textwidth]{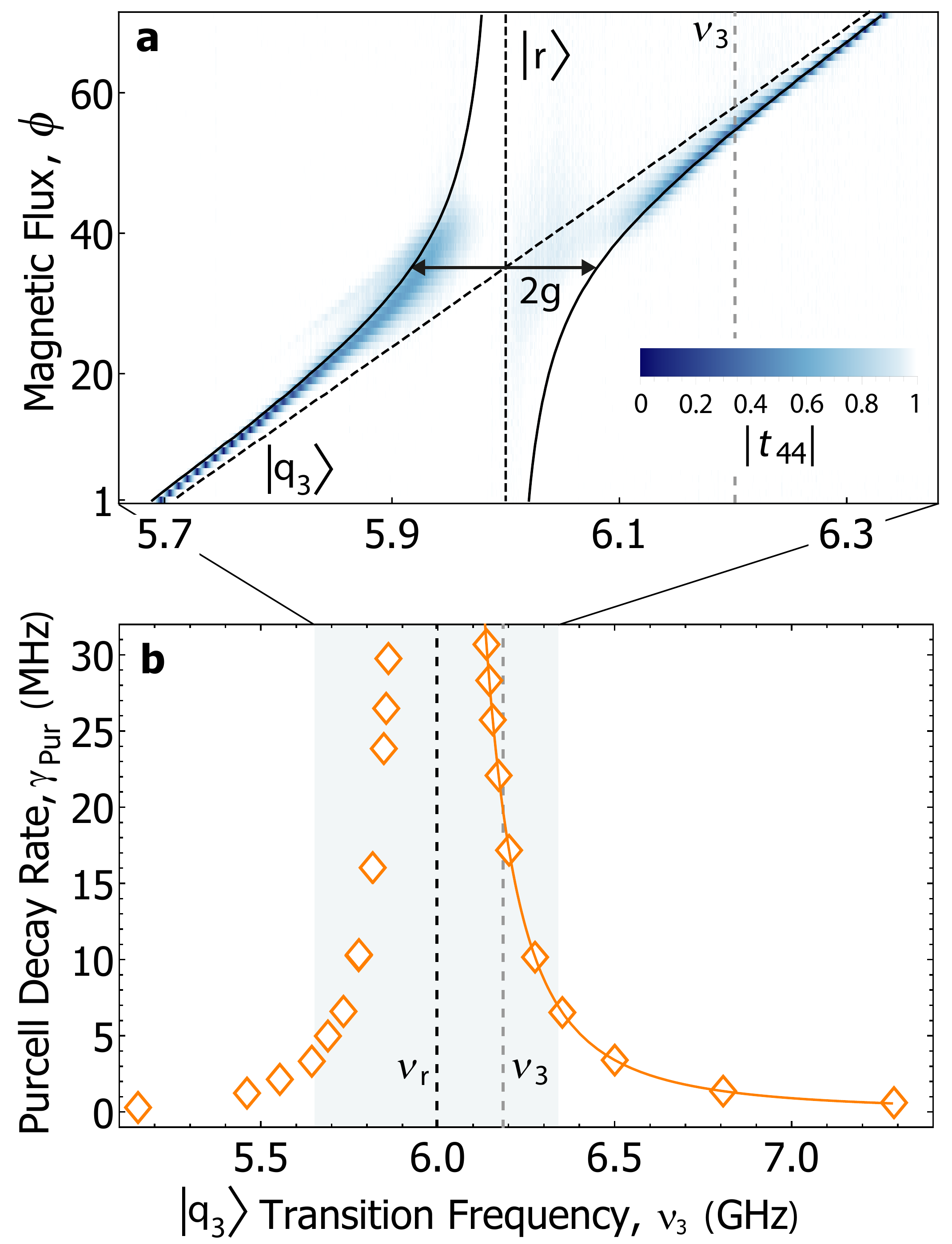} 
\end{center}
\caption{{\bf Purcell decay of $\Qthree$.}
\A~Frequency dependent reflection coefficient $\vert t_{44}\vert$ as a function of magnetic flux where the $\qthree$ transition frequency $\omega_3$ is swept linearly and the resonator fundamental frequency $\omega_{\rm{r}}$ is fixed (displayed as black dashed lines). The gray dashed line corresponds to the $\qthree$ transition frequency $\omega_\mathrm{3}/2\pi = 6.198$~GHz shown in Fig.~\ref{fig-2}b. \B~$\Qthree$ Purcell decay rate $\gamma_\mathrm{Pur}$ as a function of $\Qthree$ transition frequency extracted from the reflection coefficient measurements $\vert t_{44}\vert$. The resonator frequency $\omega_\mathrm{r}/2\pi = 6.00$~GHz is indicated with a black dashed line. The solid line is a fit to Eq.~(\ref{eq:Purcell}). The light blue area indicates the frequency range in \A~and the vertical gray dashed line indicates the $\qthree$ transition frequency, similar as in \A.}
\label{fig:Purcell}
\end{figure}

We tune the $\Qthree$ decay rate by adjusting the $\qthree$ transition frequency detuning from the extraction resonator frequency. When the resonator decay rate is large, the radiative decay rate of the qubit is enhanced by the Purcell effect \cite{Purcell1946,Houck2008}. The Purcell decay rate \cite{Sete2014}
\begin{equation}
\gamma_\mathrm{Pur} = \frac{\kappa}{2}-\frac{\sqrt{2}}{2} \sqrt{-A+\sqrt{A^2+\left(\kappa \Delta_\mathrm{3r}\right)^2}},
\label{eq:Purcell}
\end{equation}
with $A = \Delta_\mathrm{3r}^2 +4g^2 -\kappa^2/4$, depends on the coupling $g$ between the qubit and the resonator, the resonator decay rate $\kappa$ and the detuning $\Delta_\mathrm{3r}$ between the qubit and the resonator. 
In the dispersive limit ($\Delta_\mathrm{3r} \gg g$), Eq.~(\ref{eq:Purcell}) reduces to the well known expression $\gamma_\mathrm{Pur} = \kappa \left(g/\Delta_\mathrm{3r}\right)^2$ \cite{Koch2007,Sete2014}. The Purcell broadening of the $\Qthree$ spectral line is observed in a reflection coefficient measurement $\vert t_{44}\vert$ at the resonator port 4 (Fig.~\ref{fig:Purcell}a). Fitting the spectral lineshape using master equation simulations we extract the Purcell decay rate $\gamma_\mathrm{Pur}$ which is a function of the tunable $\qthree$ frequency near the resonator fundamental mode $\omega_\mathrm{r}/2\pi = 6.00$~GHz (Fig.~\ref{fig:Purcell}b). 
The extracted values $\gamma_\mathrm{Pur}$ for positive detunings above the resonator frequency agree with  Eq.~(\ref{eq:Purcell}) with coupling $g/2\pi = 90$~MHz and resonator decay rate $\kappa/2\pi = 110$~MHz. 
For $\omega_3/2\pi = 6.198$~GHz used in the presented experiments (indicated by the gray dashed line in Fig.~\ref{fig:Purcell}b) the Purcell decay rate is $\gamma_\mathrm{Pur}/2\pi \approx 20$ MHz.

\section{Rabi Rate and PSD Calibration}
\label{sec:RabiCalibration}

\begin{figure}
\begin{center}
\includegraphics[trim={0cm 0cm 0cm 0cm},clip,width=0.48\textwidth]{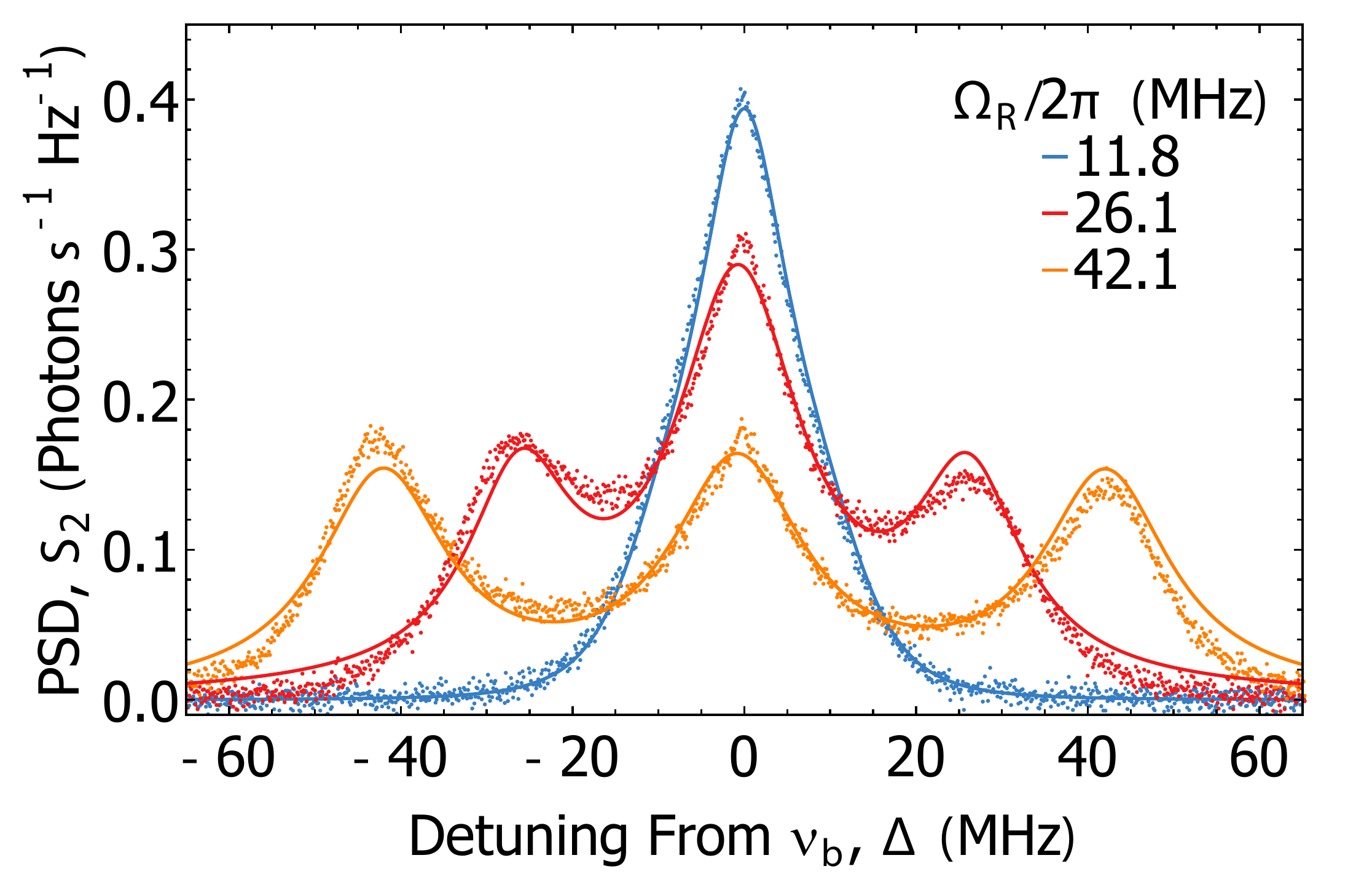} 
\end{center}
\caption{{\bf Mollow triplet of the bright state.}
Measured power spectral density (PSD) $S_2(\omega)$ of the bright state resonance fluorescence emission spectrum at the output of the waveguide for indicated coherent microwave amplitudes given in terms of the extracted Rabi rate $\Omega_\mathrm{R}$. The microwave drive frequency was set to the bright state transition frequency $\omega_\mathrm{b}$. Solid lines are fits to the Mollow triplet spectrum  \cite{Carmichael2002} for $\Omega_\mathrm{R}/2\pi = 11.8$~MHz and to the sum of three Lorentzian functions for $\Omega_\mathrm{R}/2\pi = 26.1$ and 42.1~MHz.}
\label{fig:PowerCalibration}
\end{figure}

The Rabi rate $\Omega_\mathrm{R}$ of the coherently driven bright $\bright$ state was determined from measurements of bright state resonance fluorescence power spectra $S_2(\omega)$ (Fig.~\ref{fig:PowerCalibration}). To obtain the Rabi rate $\Omega_\mathrm{R}/2\pi = 14$~MHz for the microwave powers used in our experiments we fit the resonance fluorescence spectrum to the Mollow triplet expression \cite{Carmichael2002} assuming negligible pure dephasing and non-radiative decay. For larger applied microwave powers, the full Mollow triplet \cite{Mollow1969} of the bright state emerges with resolved side peaks. We determine the Rabi rate $\Omega_\mathrm{R}$ for these amplitudes from the frequency splitting between the central and the side peaks when fitting the spectra with three Lorentzian lines. Fits to the data are shown with solid lines in Fig.~\ref{fig:PowerCalibration}.

In order to calibrate the amplitude of the measured power spectral densities we performed the same power spectrum measurement on $\Qtwo$ with all other qubits detuned by more than 1.5~GHz. For strong coherent drive, when the qubit is saturated, and assuming that pure dephasing and non-radiative decay rates are negligible the integrated power of the measured Mollow triplet is equal to $\gamma_1/2$ photons per unit time. This allows us to express the magnitude of measured PSD in Photons $\mathrm{s}^{-1} \ \mathrm{Hz}^{-1}$.

\section{Noise Generation}
\label{sec:Noise}
Low frequency noise is generated from a filtered random number time series consisting of $16\cdot10^6$ values. The bandwidth of the generated noise spans from 75 Hz to 600 MHz. The time series with a desired power spectral density $S(\omega)$ is constructed by first calculating the Gaussian random number sequence with a unit power spectral density $S(\omega) = 1$ and then applying a finite impulse response filter (FIR) with the frequency response function $H(\omega)$. We compensate AWG signal discritization distortions by pre-equalizing the digital noise series with an additional filter, constructed from the AWG output spectrum of an ideal white noise digital signal measured using a spectrum analyzer.

In this work we consider two distinct power spectral densities: (i) white noise with an exponential cutoff based on the Fermi-Dirac distribution
\begin{equation}
S_\mathrm{W}(\omega) = \frac{A_\mathrm{W}}{1+e^{\frac{\omega-\omega_\mathrm{c}}{\Delta\omega}}},\label{eq:WhiteNoise}
\end{equation}
where $A_\mathrm{W}$ is the amplitude of the function constant up to an exponential cutoff at $\omega_\mathrm{c}/2\pi = 325$~MHz with a characteristic width $\Delta\omega/2\pi = 5.44$~MHz. (ii) Noise with a Lorentzian power spectral density:
\begin{equation}
S_\mathrm{L}(\omega) = \frac{A_\mathrm{L}}{1+\left(\frac{\omega-\omega_\mathrm{L}}{\Delta\omega_\mathrm{L}/2}\right)^2},\label{eq:LorentzianNoise}
\end{equation}
where $A_\mathrm{L}$ is the amplitude, $\omega_\mathrm{L}/2\pi = 0-300$~MHz is the variable center frequency and $\Delta\omega_\mathrm{L}/2\pi = 10$~MHz is the full width at half maximum. The noise power spectral densities in Fig.~\ref{fig-2}c in the main text are measured with a spectrum analyzer at the output of the AWG.

\section{Pure Dephasing Rate Calibration}
\label{sec:DephasingCalibration}

\begin{figure}[t]
\begin{center}
\includegraphics[trim={0cm 0cm 0cm 0cm},clip,width=0.42\textwidth]{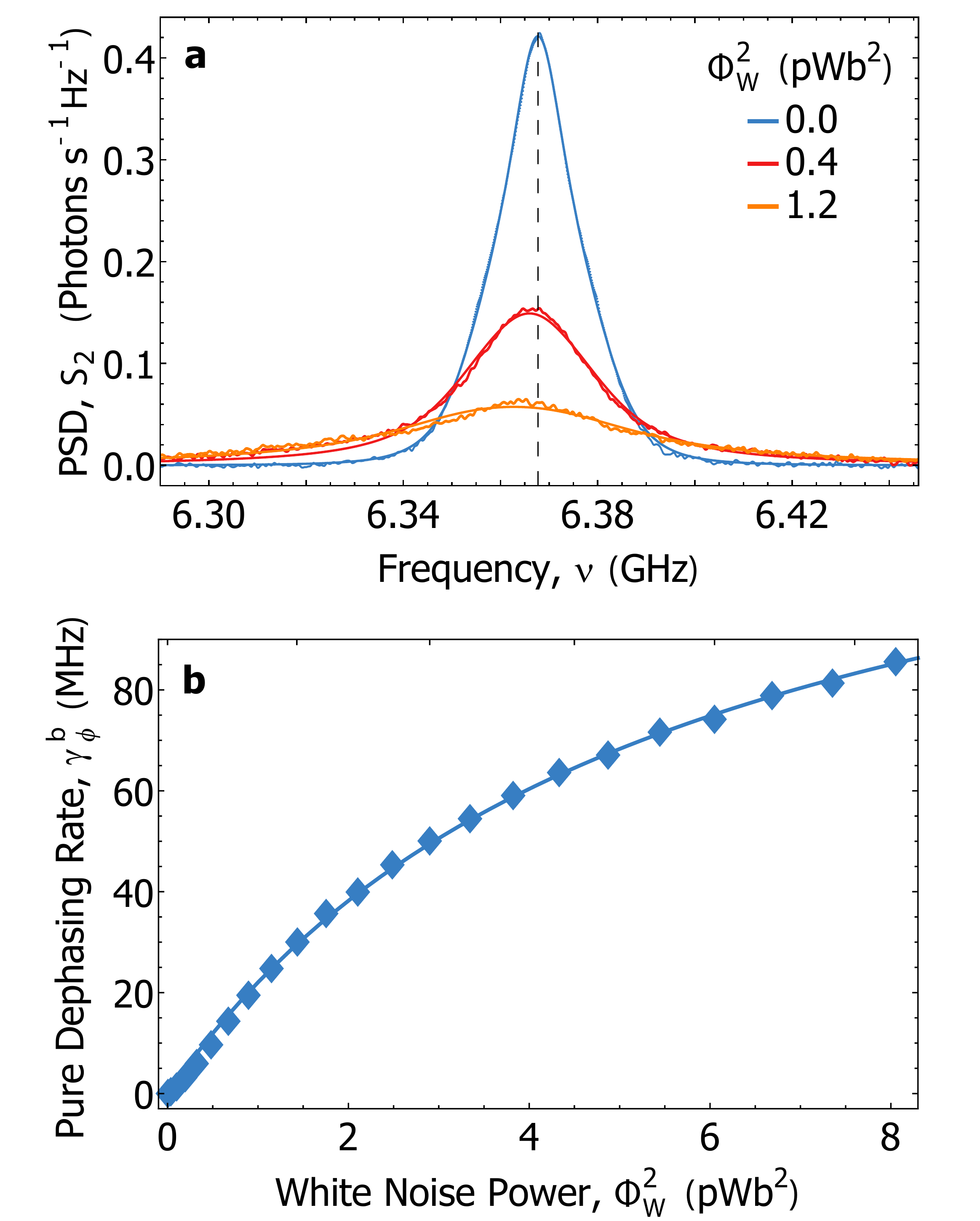} 
\end{center}
\caption{{\bf Bright state dephasing rate.}
\A~Measured power spectral density $S_2(\omega)$ of the bright mode at port 2 of the waveguide for indicated environmental applied white noise powers $\Phi_\mathrm{W}^2$. Solid lines are fits to the Mollow-triplet spectrum \cite{Carmichael2002}, see text. \B~Pure dephasing rate $\gamma_\phi^\mathrm{b}$ of the bright mode as a function of white noise power $\Phi_\mathrm{W}^2$ obtained from the fitted bright state resonance fluorescence spectra. The solid curve is a numerical calculation of the pure dephasing rate $\gamma_\phi^\mathrm{b}$ for the white noise power spectral density with a finite frequency cutoff at 325~MHz (see Fig.~\ref{fig-2}c in the main text).}
\label{fig:NoiseCalibration}
\end{figure}

White noise applied to $\Qtwo$ increases the pure dephasing rate of $\qtwo$ and consequently that of bright $\bright$ and dark $\dark$ states. The bright state pure dephasing rate $\gamma_\phi^\mathrm{b}$ is determined from the resonance fluorescence power spectrum of the bright mode measured through the waveguide $S_2(\omega)$ for indicated applied noise powers $\Phi_\mathrm{W}^2$ (see Fig.~\ref{fig:NoiseCalibration}a). Measured spectra are fitted to the Mollow triplet expression \cite{Carmichael2002} with the pure dephasing rate $\gamma_\phi^\mathrm{b}$ as a free parameter and fixed center frequency $\omega_0$, Rabi rate $\Omega_\mathrm{R}$ and decay rate $\gamma_\mathrm{b}$.
The extracted pure dephasing rate $\gamma_\phi^\mathrm{b}$ (see Fig.~\ref{fig:NoiseCalibration}b) shows an initial linear increase with applied white noise power $\Phi_\mathrm{W}^2$  as expected for ideal Markovian white power spectral density PSD \cite{Martinis2003}. Deviations from the linear dependence at higher noise powers originate from the finite cutoff of the engineered white noise (see Fig.~\ref{fig-2}c in the main text). The numerically calculated pure dephasing rate \cite{Martinis2003} using a noise power spectral density with a finite cutoff is in excellent agreement with the data (solid line in Fig.~\ref{fig:NoiseCalibration}b) indicating that the origin of the deviation from the linear dependence is the finite bandwidth of the engineered white noise.

\section{White Noise PSD Analysis}
\label{sec:WhiteNoise}

\begin{figure}
\begin{center}
\includegraphics[trim={0cm 0cm 0cm 0cm},clip,width=0.5\textwidth]{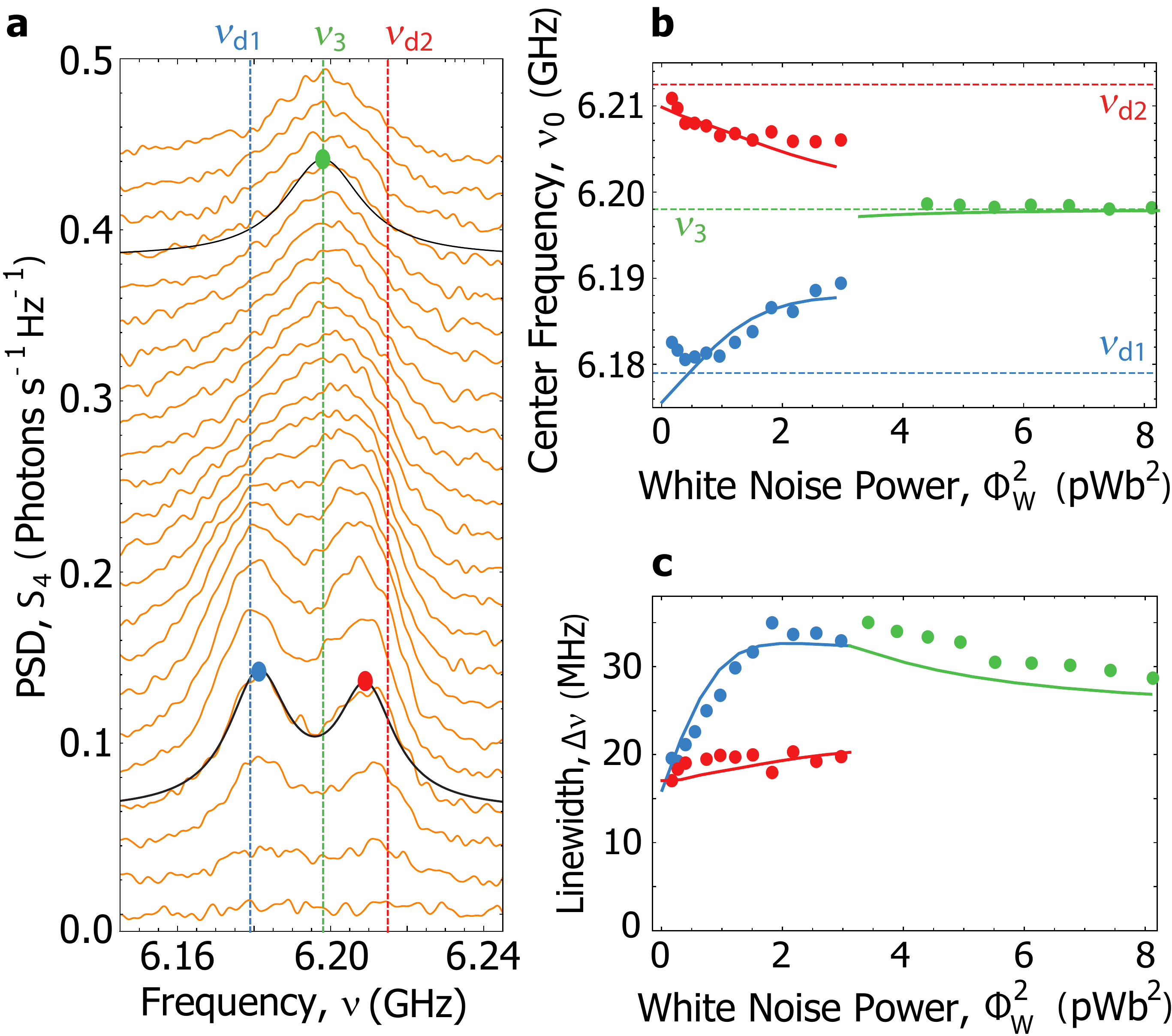} 
\end{center}
\caption{{\bf Spectra of $\done$ and $\dtwo$ dark states.}
\A~Power spectral density $S_4(\omega)$ at the resonator output port 4 for applied environmental white noise powers $\Phi_\mathrm{W}^2$ ranging from 0 to 8.1~pWb$^2$. Black solid curves show two selected fits to the data using two Lorentzian functions ($\Phi_\mathrm{W}^2 = 0.6$~pWb$^2$) and a single Lorentzian function ($\Phi_\mathrm{W}^2 = 6.1$~pWb$^2$). Vertical dashed lines indicate spectroscopically determined transition frequencies of $\done$ ($\nu_\mathrm{d1} = 6.179$~GHz), $\dtwo$ ($\nu_\mathrm{d2} = 6.216$~GHz) and $\qthree$ ($\nu_\mathrm{3} = 6.198$~GHz) states. \B~Lorentzian center frequencies as a function of white noise power $\Phi_\mathrm{W}^2$ from the fits (red, blue and green dots). Horizontal dashed lines mark $\done$, $\dtwo$ and $\qthree$ transition frequencies as in \A. \C~Lorentzian full width at half maximum (fwhm) as a function of white noise power $\Phi_\mathrm{W}^2$ determined from fitting the power spectral densities indicated in \A. Solid lines in {\B} and {\C} are results from master equation simulations.}
\label{fig:WhiteNoiseCoherentWaterfallRes}
\end{figure}

We analyze power spectral densities $S_4(\omega)$ measured at the resonator as a function of applied environmental white noise power (see Fig.~\ref{fig-3}a and Fig.~\ref{fig:WhiteNoiseCoherentWaterfallRes}a) by fitting the spectra with a sum of two Lorentzian functions 
\begin{equation}
F(\nu) = \frac{a_1}{1+4\left(\frac{\nu-\nu_{01}}{\Delta\nu_1}\right)^2} + \frac{a_2}{1+4\left(\frac{\nu-\nu_{20}}{\Delta\nu_2}\right)^2}.
\end{equation}
Here $a_i$ is the amplitude, $\nu_{0i}$ the center frequency and $\Delta\nu_i$ the full width at half maximum of the $i$-th Lorentzian. 
For higher noise powers, for which the two peaks are not resolved anymore, we fit the data to a single Lorentzian. The two resonances centered near the spectroscopically determined transition frequencies $\nu_\mathrm{d1}$ and $\nu_\mathrm{d1}$ for low white noise powers gradually shift towards the bare $\qthree$ transition frequency $\nu_3$ (Fig.~\ref{fig:WhiteNoiseCoherentWaterfallRes}b). 
At the same time the linewidth of the lower frequency resonance corresponding to $\done$ significantly broadens while the one corresponding to $\dtwo$ remains unchanged (see Fig.~\ref{fig:WhiteNoiseCoherentWaterfallRes}c).
The different dependence of the $\done$ and $\dtwo$ resonances on the applied noise is attributed to an imperfect hybridization of the $\dark$ and $\qthree$ states.
From the analyzed data we determine the cross-over from the strong to the weak coupling regime to occur at $\Phi_\mathrm{W}^2=3.0 \pm 1.0$~$\mathrm{pWb}^2$ corresponding to $\gamma_\phi^\mathrm{b}/2\pi = 50\pm 10$~MHz. This value is comparable to the $2J_\mathrm{d3}/2\pi = 37$~MHz, which is in agreement with the crossover from strong to weak coupling as discussed in the main text.

\section{Theoretical Description}
\label{sec:Theory}

\subsection{Master Equation}
\label{sec:TheoryME}

The unitary dynamics of the three coupled qubits where $\Qthree$ is coupled to the extraction resonator and $\Qone$ and $\Qtwo$ are driven by a coherent tone applied to the waveguide is described by the Hamiltonian in the rotating wave approximation
\begin{align}
\hat{\mathcal{H}} / \hbar = & \sum_{j=1}^3 \left[ \frac{(\omega_{j}-\omega_\mathrm{in})}{2}\hat{\sigma}_{j}^z + \sum_{k<j} J_{kj}\left(\hat{\sigma}_k^+\hat{\sigma}_j^- + \hat{\sigma}_j^+ \hat{\sigma}_k^-\right) \right] \nonumber \\
& +(\omega_\mathrm{r}-\omega_\mathrm{in})\hat{a}^\dag\hat{a} +
g_{3}\left(\hat{a}^\dag\hat{\sigma}_3^- +\hat{\sigma}_3^+\hat{a}\right) \label{eq:Hamiltonian}\\
&+ \frac{\Omega_\mathrm{R1}}{2} (\hat{\sigma}_{1}^++\hat{\sigma}_1^-) + \frac{\Omega_\mathrm{R2}}{2} (\hat{\sigma}_{2}^++\hat{\sigma}_{2}^-),\nonumber
\end{align}
where $\hat{\sigma}_j^z$, $\hat{\sigma}_j^+ = (\hat{\sigma}_j^x + i\hat{\sigma}_j^y)/2$ and $\hat{\sigma}_j^- = (\hat{\sigma}_j^x - i\hat{\sigma}_j^y)/2$ are Pauli operators, $\hat{a}\,(\hat{a}^\dagger)$ is the annihilation (creation) operator of the resonator's harmonic mode, $\omega_\mathrm{in}$ is the input microwave frequency and also the frequency of the reference frame. The applied low frequency noise to $\Qtwo$ is modeled as a time dependence of $\omega_2 \equiv \omega_2(t)$ (see App.~\ref{sec:EffectiveQubitPhononCoupling}). $\Omega_{\mathrm{R}j}$ is the Rabi frequency for $j$-th qubit. In the case of ideal hybridization between $\Qone$ and $\Qtwo$ ($\omega_1 = \omega_2$) the Hamiltonian can be written in the bright and dark state bases as 
\begin{align}
\hat{\mathcal{H}} / \hbar = & \sum_{j=\mathrm{b,d,3}}  (\omega_{j}-\omega_\mathrm{in})\hat{\sigma}_{j}^+\hat{\sigma}_{j}^- +(\omega_\mathrm{r}-\omega_\mathrm{in})\hat{a}^\dag\hat{a} \nonumber \\ 
&+J_\mathrm{b3}\left(\hat{\sigma}_\mathrm{b}^+\hat{\sigma}_3^- + \hat{\sigma}_3^+ \hat{\sigma}_\mathrm{b}^-\right) - J_\mathrm{d3}\left(\hat{\sigma}_\mathrm{d}^+\hat{\sigma}_3^- + \hat{\sigma}_3^+ \hat{\sigma}_\mathrm{d}^-\right)  \\
&  +
g\left(\hat{a}^\dag\hat{\sigma}_3^- +\hat{\sigma}_3^+\hat{a}\right) + \frac{\Omega_\mathrm{R}}{2} (\hat{\sigma}_\mathrm{b}^++\hat{\sigma}_\mathrm{b}^-) ,\nonumber
\end{align}
where $\hat{\sigma}_\mathrm{b}^\pm = (\hat{\sigma}_1^\pm + \hat{\sigma}_2^\pm)/\sqrt{2}$ and $\hat{\sigma}_\mathrm{d}^\pm = (\hat{\sigma}_1^\pm - \hat{\sigma}_2^\pm)/\sqrt{2}$ are bright and dark state creation and annihilation operators,  $\omega_\mathrm{b} = \omega_\mathrm{1}+J_{12}$, $\omega_\mathrm{d} = \omega_\mathrm{1}-J_{12}$ are bright and dark state transition frequencies, $J_\mathrm{b3} = (J_{23}+J_{13})/\sqrt{2}$ and $J_\mathrm{d3} = (J_{23}-J_{13})/\sqrt{2}$ are coupling rates between $\Qthree$ and bright, and $\Qthree$ and dark state, respectively, and $\Omega_\mathrm{R} = \sqrt{2}\Omega_\mathrm{R1}$ is the bright state Rabi frequency.

The full dynamics including the non-unitary terms is given by the Lindblad equation
\begin{align}
\dot{\rho}=\mathcal{L}(\rho),\label{eq:Lindblad}
\end{align}
where $\rho$ is the density matrix and $\mathcal{L}(\rho)$ is given by
\begin{align}
\mathcal{L}(\rho) = &- \frac{i}{\hbar} \left[ \hat{\mathcal{H}},\rho\right] \nonumber \\
&+ \gamma_\mathrm{b}(1+n_\mathrm{th})L(\sigma_\mathrm{b}^-)\rho + \gamma_\mathrm{b}n_\mathrm{th}L(\sigma_\mathrm{b}^+)\rho
\nonumber \\
&+ \gamma_\mathrm{d}(1+n_\mathrm{th})L(\sigma_\mathrm{d}^-)\rho + \gamma_\mathrm{d}n_\mathrm{th}L(\sigma_\mathrm{d}^+)\rho \label{eq:ME}
\\
&+ \frac{\gamma_\phi}{2} L(\sigma_2^+\sigma_2^- - \sigma_2^-\sigma_2^+)\rho
\nonumber \\
&+ \kappa L(a)\rho. \nonumber
\end{align}
Here $L(\sigma)$ is the Lindblad superoperator
\begin{align}
L(\sigma)\rho = \hat{\sigma} \rho \hat{\sigma}^\dagger - \frac{1}{2}\left( \hat{\sigma}^\dagger\hat{\sigma}\rho + \rho\hat{\sigma}^\dagger\hat{\sigma} \right)
\end{align}
and $n_\mathrm{th} < 0.01$, a typical thermal occupation for our experiments.

\subsection{Simulations}
\label{sec:TheorySim}
Simulations of power spectral densities $S_2(\omega)$ at the waveguide output port 2 and the resonator $S_4(\omega)$ at port 4  as well as the integrated power at the waveguide $P_2$, the resonator $P_4$ and the transfer efficiency $\eta$ are performed with QuTiP 3.1.0 \cite{Johansson2013a}. All simulations are done using the Lindblad master equation, which is sufficient for time independent decay channels, as well as the Bloch-Redfield master equation to account for the finite noise frequency cutoff (see Fig.~\ref{fig-2}c). Both methods yield identical results confirming that the noise PSD cutoff frequency ($\approx\,$325~MHz) is high enough for the interaction between our circuit and the environment to be considered in the Markovian approximation.


Waveguide and resonator spectra are calculated for the steady state $\dot \rho = 0$ via two-time correlation functions
\begin{equation}
	s(\omega) = \int_{-\infty}^\infty {\braket{A(\tau) B(0)} e^{-i\omega \tau}}\, d\tau.
\end{equation}
As a proxy for the waveguide emission we use the bright state $\bright$ correlation function $\braket{\hat\sigma_\mathrm{b}^+(\tau) \hat\sigma^-_\mathrm{b}(0)}$ and for the resonator emission we use $\braket{\hat a^\dag(\tau) \hat a(0)}$.
The correct magnitude is achieved by multiplication with the respective radiative decay rates $\gamma_\mathrm{b}/2$ and $\kappa$, where the factor 1/2 for the $\gamma_\mathrm{b}$ reflects the detection of only half of the photons emitted into the waveguide when measured only at port 2 and omitting port 1
\begin{equation}
\begin{split}
	S_\mathrm{2}(\omega) \approx \frac{\gamma_\mathrm{b}}{2} \int_{-\infty}^\infty {\braket{\hat\sigma_\mathrm{b}^+(\tau) \hat\sigma^-_\mathrm{b}(0)} e^{-i\omega \tau}}\, d\tau,\\
	S_\mathrm{4}(\omega) = \kappa \int_{-\infty}^\infty {\braket{\hat a^\dag(\tau) \hat a(0)} e^{-i\omega \tau}}\, d\tau.
\end{split}
\end{equation}
As in the experimental analysis we obtain the full power as an integral of the power spectral density over frequency.

For the simulations shown in Fig.~\ref{fig-3}b and Figs.~\ref{fig-4}a,c,  in the main text we use system parameters specified in Table~\ref{tab:systemParameters}. 
In Figs.~\ref{fig-4}a,c of the main text the data is plotted against the bright state pure dephasing rate, which is related to the $\Qtwo$ pure dephasing rate as $\gamma_\phi^\mathrm{b} = \gamma_\phi/2$ assuming that the bright state is an equal superposition of $\qone$ and $\qtwo$.

To reproduce experimental results for incoherent excitation a Lindblad master equation was solved without the last two terms in Eq.~(\ref{eq:Hamiltonian}). A thermal occupation of $n_\mathrm{th} = 0.3$ was used to compute the integrated re-emitted $P_2$ and extracted $P_4$ powers as well as the transport efficiency $\eta$ shown with solid lines in Fig.~\ref{fig-4}c.

\begin{table}
\begin{center}
  \begin{tabular}{ | c | c | c | }
    \hline
 	Description & Parameter & Value  \\ \hline
    $\qone$ transition frequency & $\omega_1/2\pi$ & 6.277~GHz  \\
    $\qtwo$ transition frequency & $\omega_2/2\pi$ & 6.277~GHz \\
    $\qthree$ transition frequency & $\omega_3/2\pi$ & 6.161~GHz \\
    Resonator frequency & $\omega_\mathrm{r}/2\pi$ & 6.000~GHz \\ \hline
    Coupling between $\Qone$ and $\Qtwo$ & $J_{12}/2\pi$ & 83.5~MHz \\
    Coupling between $\Qtwo$ and $\Qthree$ & $J_{23}/2\pi$ & 33.4~MHz \\
    Coupling between $\Qone$ and $\Qthree$ & $J_{13}/2\pi$ & 3.67~MHz \\
    Coupling between $\Qthree$ and & $g_{3}/2\pi$ & 90~MHz \\ 
    the resonator &  &  \\ \hline
    $\bright$ state decay rate & $\gamma_\mathrm{b}/2\pi$ & 12.4~MHz \\
    Resonator decay rate & $\kappa/2\pi$ & 110~MHz \\ \hline
    Bright state Rabi rate & $\Omega_\mathrm{R}/2\pi$ & 14.0~MHz \\
    Input field frequency & $\omega_\mathrm{in}/2\pi$ & 6.368~GHz \\
    \hline
  \end{tabular}
\end{center}
\caption{System parameters used for simulations. All parameters are experimentally determined from spectroscopic measurements except $g$ and $\kappa$ which were adjusted within their experimental uncertainty. }
\label{tab:systemParameters}
\end{table}

\subsection{Rate Equations}
\label{sec:TheoryRE}
The rate equations for the populations of the bright $\bright$ state $p_\mathrm{b} = \rho_\mathrm{bb}$, dark $\dark$ state $p_\mathrm{d} = \rho_\mathrm{dd}$ and $\qthree$ state $p_\mathrm{3} = \rho_\mathrm{33}$ are derived from the Lindblad equation of motion [Eq.~(\ref{eq:Lindblad})] where the coupling of $\Qthree$ to the resonator is approximated by an effective Purcell decay.
The incoherent dephasing $\gamma_\phi/2\, L[\sigma_{2}^z]\rho$, with $\gamma_\phi = 2\gamma_\phi^\mathrm{b}$ leads to a decay of all coherences involving $\Qtwo$
\[\dot \rho_{i2} = -\gamma_\phi \rho_{i2} \quad \forall i \neq 2.\]
In the bright/dark state basis this operator describes incoherent transport $\gamma_\phi^\mathrm{b} (L[\sigma_\mathrm{b}^+\sigma_\mathrm{d}^-]+L[\sigma_\mathrm{d}^+\sigma_\mathrm{b}^-])\rho$ between $\dark$ and $\bright$ towards an equilibrium population determined by
\[\frac{d}{dt}(\rho_\mathrm{bb}-\rho_\mathrm{dd}) = -2\underbrace{\gamma_\phi^\mathrm{b}}_{k_\mathrm{bd}} (\rho_\mathrm{bb}-\rho_\mathrm{dd}),\]
The bright state population $	\dot \rho_\mathrm{bb} \propto \Omega_\mathrm{R} \mathrm{Im}(\rho_\mathrm{gb}) $
is reduced by incoherent dephasing since the coherence between the ground and the bright state $\rho_\mathrm{gb}$ evolves as
\begin{equation}
	\dot \rho_\mathrm{gb} \propto i\frac{\Omega_\mathrm{R}}{2} (\rho_\mathrm{gg}-\rho_\mathrm{bb}) - \frac{\gamma_\mathrm{b}+2\gamma_\phi^\mathrm{b}}{2} \rho_\mathrm{gb}.\nonumber
\end{equation}
Thus absorption is reduced by dephasing when $\gamma_\phi^\mathrm{b} \gtrsim \Omega_\mathrm{R}$. In the steady state the bright state population can be written as
\begin{equation}
\dot \rho_\mathrm{bb} \propto \frac{\Omega_\mathrm{R}^2}{\gamma_\mathrm{b}+2\gamma_\phi^\mathrm{b}} (\rho_\mathrm{gg}-\rho_\mathrm{bb}).\nonumber
\end{equation}
The coherent population transfer from $\dark$ to $\qthree$ is defined by their coherence
\[ \frac{d}{dt}(\rho_\mathrm{dd}-\rho_{33}) \propto -4 J_\mathrm{d3} \mathrm{Im}(\rho_\mathrm{d3}),\]
which in turn is controlled by the coherent coupling and the incoherent dephasing
\begin{equation}
\dot \rho_\mathrm{d3} \propto -i J_\mathrm{d3}(\rho_{33}-\rho_\mathrm{dd}+\rho_\mathrm{db}) -\frac{\gamma_\mathrm{Pur}+2\gamma_\phi^\mathrm{b}}{2} \rho_\mathrm{d3}. \nonumber
\end{equation}
The scale at which coherent transport is expected to be reduced due to noise is  $\gamma_\phi^\mathrm{b} \gtrsim J_\mathrm{d3}$. 
To derive the F\"orster transport rates $k_\mathrm{gb}$ between the ground $\ket{\mathrm{g}}$ and the bright state $\bright$ and the dark $\dark$ and $\qthree$ state $k_\mathrm{d3}$ in the strong dephasing limit we assume that coherences are small, if they are not participating in transport, and that derivatives of coherences are negligible \cite{Zhang2017f}. 
For $\rho_{db}\approx 0$ and $\dot\rho_{d3}=\dot \rho_{gb}=0$ we therefore have
\begin{align}
	\rho_\mathrm{d3} &=  i \frac{-J_\mathrm{d3}}{\gamma_\mathrm{Pur}/2+ \gamma_\phi^\mathrm{b}}(\rho_{33}-\rho_\mathrm{dd}),\nonumber\\
    \rho_\mathrm{gb} &=  i \frac{\Omega_\mathrm{R}}{\gamma_\mathrm{b}+ 2\gamma_\phi^\mathrm{b}}(\rho_\mathrm{gg}-\rho_\mathrm{bb})\label{eq-rates}.
\end{align}
With the help of the above expressions we find the rate equations
\begin{align}
	\dot p_\mathrm{g} &= -k_\mathrm{gb} (p_\mathrm{g}-p_\mathrm{b}) +\gamma_\mathrm{b}p_\mathrm{b} +\gamma_\mathrm{Pur}p_\mathrm{3}, \nonumber\\
	\dot p_\mathrm{b} &= k_\mathrm{gb} (p_\mathrm{g}-p_\mathrm{b}) - k_\mathrm{bd}  (p_\mathrm{b}-p_\mathrm{d}) -\gamma_\mathrm{b}p_\mathrm{b}, \label{eq-re}\\
	\dot p_\mathrm{d} &= k_\mathrm{bd} (p_\mathrm{b}-p_\mathrm{d}) -k_\mathrm{d3}(p_\mathrm{d}-p_{3}),\nonumber\\
    \dot p_{3} &= k_\mathrm{d3} (p_\mathrm{d}-p_{3}) - \gamma_\mathrm{Pur}p_{3},\nonumber
\end{align}
for $\ket{\mathrm{g}}$, $\bright, \dark\, \mathrm{and}\, \qthree$ populations with transfer rates
\begin{align}
	k_\mathrm{gb} &= \frac{\Omega _\mathrm{R}^2}{\gamma_\mathrm{b}+2\gamma _{\phi}^\mathrm{b}}, \nonumber\\
    k_\mathrm{d3} &= \frac{4J_\mathrm{d3}^2}{\gamma_\mathrm{Pur}+2\gamma_\phi^\mathrm{b}},\nonumber \\
    k_\mathrm{bd} &= \gamma_\phi^\mathrm{b},\nonumber
\end{align}
where state populations are bound by $p_i\in[0,\,1]$.
Since $\gamma_\mathrm{b} \approx \gamma_\mathrm{Pur}$ and $\Omega_\mathrm{R} \ll 2J_\mathrm{d3}$, the reduction of absorption for increasing $\gamma_\phi^\mathrm{b}$ happens before the complete decoupling of $\dark$ and $\qthree$. From Eq.~(\ref{eq-rates}) we see that population transfer between $\dark$ and $\qthree$ is always coherent although $\dark$ is populated incoherently and the transfer is suppressed by the dephasing rate $\gamma_\phi^\mathrm{b}$.

The transfer efficiency is calculated using steady state solutions of Eqs.~(\ref{eq-re}) as
\begin{equation}
\eta  = \frac{\gamma_\mathrm{Pur} p_{3}}{\gamma_\mathrm{Pur} p_\mathrm{3} + \gamma_\mathrm{b} p_\mathrm{b}},
\end{equation}
where we assumed $p_g = 1$. The efficiency has a maximum at $\gamma_\phi^\mathrm{b}  = \sqrt{2}J_\mathrm{d3}\approx J_{23}$ where it can be expressed as
\begin{align}
\eta  = \frac{1}{1+ \sqrt{2}\frac{\gamma_\mathrm{b}}{J_\mathrm{d3}} + \frac{\gamma_\mathrm{b}}{\gamma_\mathrm{Pur}}  + \frac{\gamma_\mathrm{b}\gamma_\mathrm{Pur}}{4J_\mathrm{d3}^2}}.
\end{align}
Assuming that $2J_\mathrm{d3} \gg \gamma_\mathrm{b},\gamma_\mathrm{Pur}$, the efficiency can be written as
\begin{eqnarray}
\eta \approx (1-\gamma_\mathrm{b}/\gamma_\mathrm{Pur}),
\end{eqnarray}
as stated in the main text.

\begin{figure*}
\includegraphics[width=1.00\textwidth]{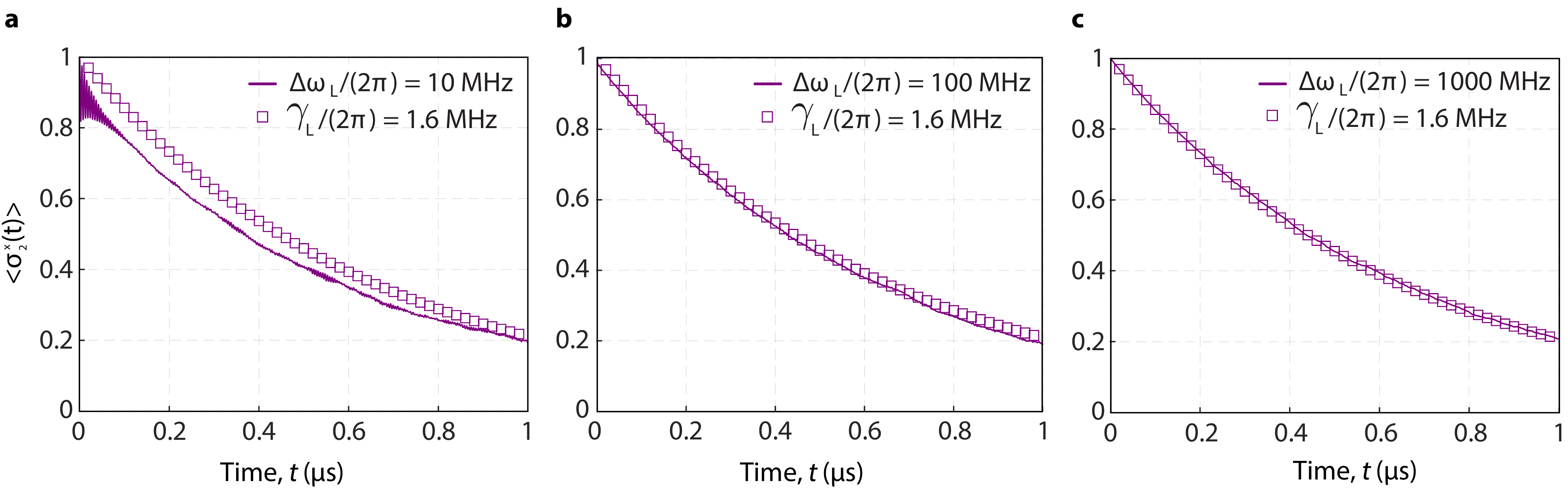}
\caption{{\bf Dephasing induced by Lorentizan noise.}
Dynamics of $\avg{\sigma_2^x(t)}$ as a function of Lorentzian noise power bandwidth $\Delta\omega_{\rm L}/2\pi =$~\{10, 100, 1000\}~MHz. The noise amplitude $A$ is decreased with increasing $\Delta\omega_{\rm L}$ so as to keep $\gamma_{\rm L}$ fixed at $\gamma_{\rm L}/2\pi = 1.6~{\rm MHz}$. Open squares indicate a calculated exponential decay of $\avg{\sigma_2^x(t)}$ within the Markov approximation [see Eq.~(\ref{lorentzDeph})] for $\gamma_{\rm L}/2\pi = 1.6~{\rm MHz}$. As the Lorentzian noise correlation time $1/\Delta\omega_\mathrm{L}$ decreases (left to right), the system dynamics are increasingly more faithfully described within the Markov approximation [Eq.~(\ref{lorentzDeph})]. }
\label{fig:sigmaXDyn}
\end{figure*}

\subsection{Dephasing Rate due to Lorentzian Noise}
\label{sec:TheoryLorNoise}

Flux noise acting on ${\rm Q}_2$ leads to an effective dephasing of the qubit with a dephasing rate depending on the noise power spectrum. In this section, we consider the effect of Lorentzian flux noise, in particular how it differs from white flux noise, and the emergence of non-Markovian effects. We study the essential physics with a single qubit model for ${\rm Q}_2$ alone, described by the Hamiltonian
\begin{align}
\mathcal{H}_2 = \frac{\omega_2}{2}\sigma_2^z + \xi(t)\sigma_2^z.\label{eq:Ham2}
\end{align}
Here, $\omega_2$ is the qubit transition frequency and $\xi(t)$ represents the input flux noise with power spectral density $S_{\xi\xi}[\omega]$,
\begin{align}
\avg{\xi(t)} = 0~,~S_{\xi\xi}[\omega] = \int_{-\infty}^{\infty} d\tau~e^{-i\omega \tau} \avg{\xi(\tau)\xi(0)}
\end{align}
In the experiment, $S_{\xi\xi}[\omega]$ is either a flat spectrum up to a certain cutoff frequency $S_\mathrm{W}(\omega)$ [Eq.~(\ref{eq:WhiteNoise})] or a Lorentzian spectrum $S_\mathrm{L}[\omega]$ [Eq.~(\ref{eq:LorentzianNoise})].


The Hamiltonian [Eq.~(\ref{eq:Ham2})] is non-demolition with respect to $\sigma_2^z$ since $[\mathcal{H}_2,\sigma_2^z] = 0$, but leads to an effective decay of $\avg{\sigma_2^x}$ and $\avg{\sigma_2^y}$ components, which we derive next. The effective dephasing rate is extracted by studying the dynamics of these observables. Moving into a frame rotating at $\omega_2$, we write the Heisenberg equations of motion for these operators as
\begin{align}
\dot{\sigma}_2^x &= -2\xi(t) \sigma_2^y, \nonumber \\
\dot{\sigma}_2^y &= +2\xi(t) \sigma_2^x.
\end{align} 
The above coupled system is formally solved to obtain a single dynamical equation for $\sigma_2^x(t)$. Averaging this equation under noise realizations, we arrive at
\begin{align}
\dot{\sigma}_2^x(t) = -4\int_0^td\tau~\avg{\xi(t)\xi(t-\tau)}\sigma_2^x(t-\tau).
\end{align}
Note that the noise autocorrelation appears in the above memory kernel; it is then possible to proceed via a Markovian approximation \textit{provided} the noise correlations decay much faster than the relaxation dynamics of the system, which are themselves driven by the noise. We make this condition precise in a self-consistent way. Assuming Markovian approximation, we can drop the system's dependence on its past history via the memory kernel and extend the integral's upper limit to infinity, thus obtaining
\begin{align}
\dot{\sigma}_2^x(t) = -4\sigma_2^x \int_0^{\infty} d\tau~\avg{\xi(\tau)\xi(0)}.
\end{align}
The remaining integral is simply half the zero frequency power spectral density of the noise signal (ignoring the principle part that leads to a Lamb shift contribution, not dephasing), so that
\begin{align}
\dot{\sigma}_2^x(t) = -2S_{\xi\xi}[0]\cdot\sigma_2^x.
\label{decay} 
\end{align}
Hence, the noise signal drives system decay at a rate $2S_{\xi\xi}[0]$ \textit{within} the Markovian approximation. For white noise, which is $\delta$-correlated, correlations always decay faster than the induced decay. Using $S_{\xi\xi}[0]$ for white noise, the dephasing rate $\gamma_\phi$ is given by:
\begin{align}
\gamma_\phi = 2A_\mathrm{W}
\end{align}
More interesting is the case of Lorentzian noise. Eq.~(\ref{decay}) yields a decay rate $\gamma_{\rm L}$ for this case as well, so long as the Lorentzian noise `appears' white, namely when the correlation time of the Lorentzian noise signal is much shorter than the time scale of the decay it induces, $1/\gamma_{\rm L}$. Since the Lorentzian noise correlation time is on the order of its inverse bandwidth $1/\Delta\omega_{\rm L}$, we require $1/\Delta\omega_{\rm L} \ll 1/\gamma_{\rm L}$, or $\Delta\omega_{\rm L} \gg \gamma_{\rm L}$. If this is the case, the dephasing rate is given by
\begin{align}
\gamma_{\rm L} = 2A_\mathrm{L} \frac{\left( \frac{\Delta\omega_{\rm L}}{2} \right)^2}{ \omega_{\rm L}^2 + \left(\frac{\Delta\omega_{\rm L}}{2}\right)^2 }~,~\Delta\omega_{\rm L} \gg \gamma_{\rm L}.
\label{lorentzDeph}
\end{align}
Clearly, the dephasing rate is reduced compared to the white noise value for the same noise amplitudes ($A_\mathrm{W} = A_\mathrm{L}$). This is due to the colored noise spectrum which is not in fact equal at all frequencies.

However, we caution that the above expression holds only when the Lorentzian noise bandwidth is much larger than the induced decay rate, $\Delta\omega_{\rm L} \gg \gamma_{\rm L}$. In the current experiment, this condition is not met and the Markovian approximation should not hold. To observe dynamics in this regime for the simple single qubit model, we numerically compute the noise-averaged dynamics of $\avg{\sigma_2^x}$ as governed by $\mathcal{H}_2$, over multiple realizations of the noise $\xi(t)$. By varying the bandwidth $\Delta\omega_{\rm L}$ of the Lorentzian noise, we are able to explore both Markovian and non-Markovian regimes. For each $\Delta\omega_{\rm L}$, the noise amplitude $A$ is chosen such that the decay rate within the Markovian approximation has the fixed value $\gamma_{\rm L}/2\pi = 1.6~{\rm MHz}$. 
The dynamics of $\avg{\sigma_2^x(t)}$ (solid line in Fig.~\ref{fig:sigmaXDyn}) for $\Delta\omega_{\rm L}/2\pi= \{10,100,1000\}~{\rm MHz}$ increasing from left to right is approximately an exponential decay. As $\Delta\omega_{\rm L}$ becomes large in comparison to $\gamma_{\rm L}$ - that is, when the noise correlation time $1/\Delta\omega_\mathrm{L}$ becomes increasingly short relative to the noise-induced system relaxation time $1/\gamma_\mathrm{L}$ - the Lorentzian noise-induced dynamics approach those predicted within the Markovian approximation (open squares in Fig.~\ref{fig:sigmaXDyn}). Due to its non-Markovian effects Eq.~(\ref{decay}) does not exactly describe the decay of coherence for the Lorentzian noise, however, it offers a meaningful estimation that is used when comparing the effects of the two types of noise.

\begin{figure}
\includegraphics[trim={0cm 0cm 0cm 0cm},clip,width=0.38\textwidth]{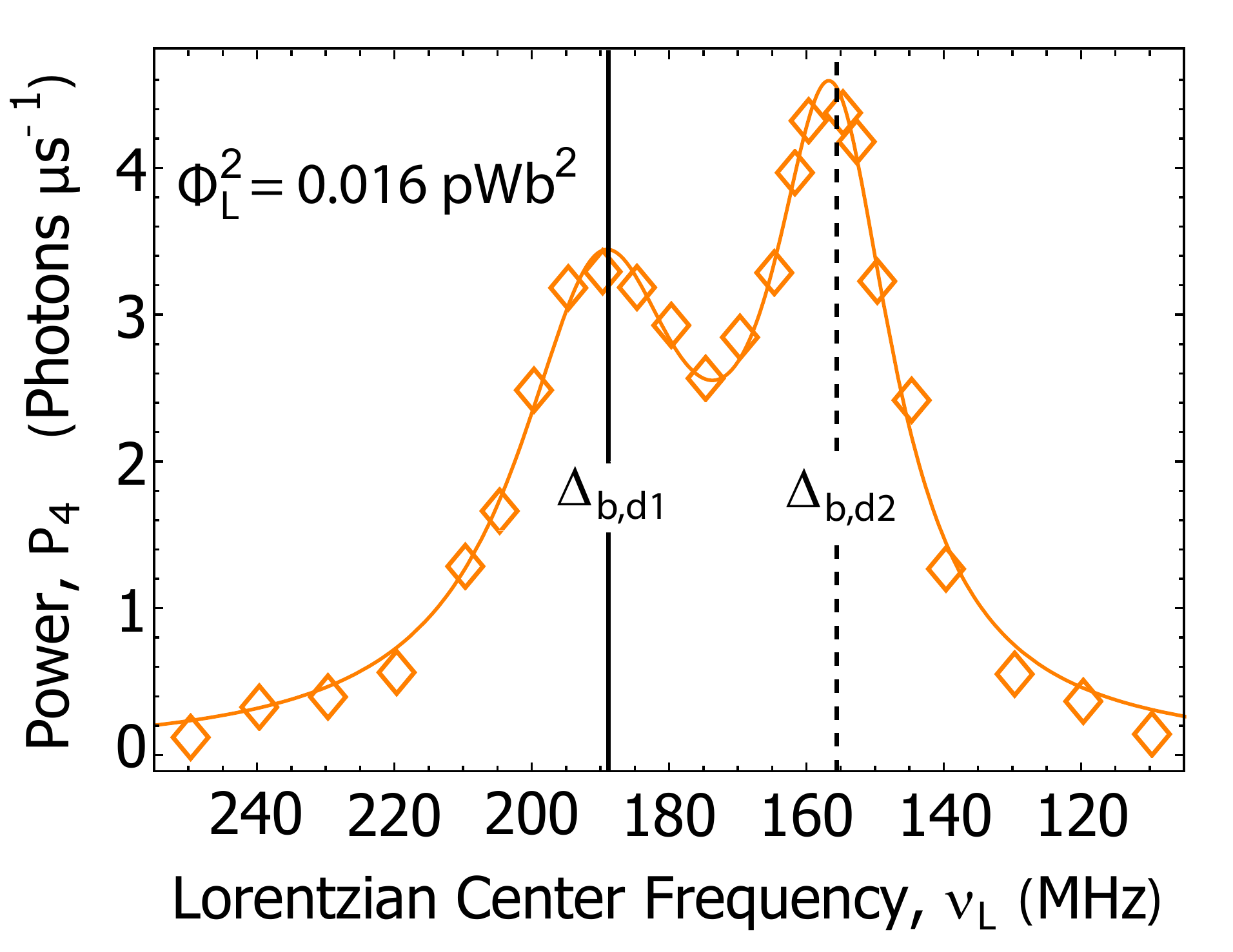} 
\caption{{\bf Extracted power $P_4$ as a function of center frequency $\nu_\mathrm{L}$ of Lorentzian noise spectrum. }
This measurement was performed at Lorentzian noise power $\Phi_\mathrm{L}^2 = 0.016$~pWb$^2$ and coherent excitation. Orange solid line is a fit to a sum of two Lorentzians. }
\label{fig:LorCentFreqSweep}
\end{figure}

\begin{figure*}
\includegraphics[width=1.01\textwidth]{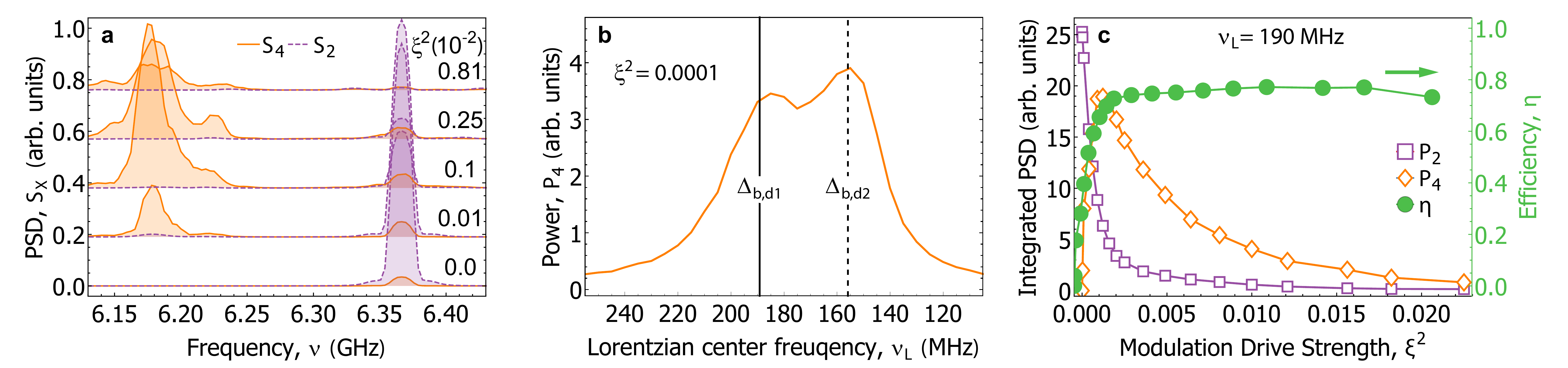}
\caption{{\bf Stochastically averaged master equation simulations.}
\A~Power spectral density calculated at the resonator $S_4(\omega)$ (orange) and waveguide $S_2(\omega)$ (purple) output for Lorentzian noise with a center frequency of $\nu_\mathrm{c} = 190$~MHz and variable modulation drive strength $\xi^2$. \B~Calculated integrated power $P_4$ as a function of the Lorentzian center frequency $\nu_\mathrm{L}$ at fixed modulation drive strength $\xi^2 = 0.0001$. \C~Integrated powers $P_2$ (purple) and $P_4$ (orange) and the transfer efficiency $\eta$ (green) as a function of Lorentzian noise modulation drive strength $\xi^2$. Power spectra at lowest noise power are matched with experiment to fix overall scaling factors, which are then used to scale all other spectra.}
\label{fig:StochasticSim}
\end{figure*}

\subsection{Simulations with Lorentzian Noise}
\label{sec:StochasticME}

Lorentzian noise spectra employed in the discussed experiments have bandwidths $\Delta\omega_{\rm L}$ that are on the order of, or slower, than the pertinent circuit decay rates. This implies that noise autocorrelation decay times surpass typical system relaxation timescales. As such, a noise environment structured in this way can give rise to non-Markovian dynamics of the system density matrix: the system's state at time $t$ can be affected by its history over a time set by the autocorrelation time of the applied noise. Note that this is the case even if the noise signal itself is entirely independent of the system evolution, as in the present setup, where the noise is algorithmically generated. Integrating out the Lorentzian noise signal yields complex memory kernels that cannot be collapsed, unlike the case for Markovian dynamics. Our approach incorporates the Lorentzian noise environment as part of the system dynamics. In this way, we may still employ a Lindblad master equation for simulations of the system density matrix, at the cost of having to deal with a stochastic term describing the system's evolution.

To proceed, we add to the system Hamiltonian in Eq.~(6) a modulation of the $\Qtwo$ energy splitting, given by the time series $\xi(t)$:
\begin{align}
\mathcal{H}_{\rm \phi} = \xi(t) \sigma_2^z \equiv \xi_0 \cos \left[ \omega_\mathrm{L} t + \phi(t) \right]\sigma_2^z 
\end{align}
Here, $\phi(t)$ is a \textit{random} variable describing phase noise, characterised by its statistical mean and variance:
\begin{align}
\avg{\phi} = 0 ~,~\avg{\phi^2} = \Delta\omega_\mathrm{L} t
\end{align}
The variance being linear in time indicates that the phase undergoes diffusion, with the parameter $\Delta\omega_\mathrm{L}$ characterizing the strength of this diffusion. The phase noise $\phi(t)$ is often referred to as Brownian noise or a Wiener process in other contexts and is the integral of Gaussian white noise. $\xi(t)$ has a Lorentzian power spectral density~\cite{Lax1967}
\begin{align}
S_{\xi\xi}(\omega) = \xi_0^2 \frac{ \frac{\Delta\omega_\mathrm{L}}{2} }{ (\omega-\omega_\mathrm{L})^2 + \left( \frac{\Delta\omega_\mathrm{L}}{2} \right)^2 }
\end{align}
with a constant integrated power proportional to $\xi_0^2$ (independent of the value of $\Delta\omega_\mathrm{L}$). Note that $S_{\xi\xi}(\omega)$ is written for $\omega > 0$; a symmetrical contribution exists for negative frequencies since $\xi(t)$ is a classical signal. Finally, note that taking $\Delta\omega_\mathrm{L} \to 0$ formally yields a coherent modulation of the $\Qtwo$ energy splitting at frequency $\omega_\mathrm{L}$.

For small $\xi_0$ we solve the master equation, Eq.~(\ref{eq:Lindblad}), with the addition of $\mathcal{H}_{\rm \phi}$ to the system Hamiltonian. The cost of adding a stochastic term to the system evolution is that any physical quantity must be computed via an explicit averaging procedure. In the Markovian approxiation, an equivalent procedure is implicitly carried out when `tracing out the bath'. For a given set of system parameters, we propagate the Master equation to long times to obtain an approximate steady state density matrix $\rho_{\rm ss}$. Then, steady state correlation functions are computed starting with the system in $\rho_{\rm ss}$. To obtain meaningful results, these computations are repeated over multiple realizations of $\xi(t)$; the relevant resonator and transmission line power spectra are then given respectively by:
\begin{align}
&S_{\rm 2}(\omega) = \\
&\quad\frac{\gamma_\mathrm{b}}{2}\int_{-\infty}^{\infty} d\tau e^{-i\omega\tau} \avg{ (\sigma_b^+(\tau)-\avg{\sigma_b^+}_{\rm ss})(\sigma_b^-(0)-\avg{\sigma_b^-}_{\rm ss}) }_{\phi} \nonumber  \\
&S_{\rm 4}(\omega) = \kappa \int_{-\infty}^{\infty} d\tau e^{-i\omega\tau} \avg{ a^{\dagger}(\tau)a(0) }_{\phi} 
\end{align}
where $\avg{\cdot}_{\phi}$ indicates an ensemble average over multiple realizations of $\phi(t)$. The subtraction of steady state averages from the bright state correlation function serves to remove the Rayleigh scattered peak from the transmission spectrum.

For stationary problems, a variety of methods exist to compute the above power spectra directly in the frequency domain, foregoing the need for a Fourier transform. Such techniques do not apply here, following inclusion of the explicitly time-dependent Lorentzian noise term. We find that the aperiodic, finite nature of computed correlation functions here leads to well known artifacts in their numerical Fourier transforms, namely a broad noise spectrum in the frequency domain due to spectral leakage. To suppress this noise, we apply a standard, total-power-preserving (Blackman) windowing function prior to performing the Fourier transform. This technique non-uniformly modifies the power at every frequency component, in different ways for each correlation function, restricting us to making only qualitative comparisons with experiment. 

Simulated spectra $S_4(\omega)$ and $S_2(\omega)$ as a function of increasing Lorentzian noise power (Fig.~\ref{fig:StochasticSim}a), integrated power at the resonator $P_4$ as a function of Lorentzian center frequency $\nu_\mathrm{L}$ (Fig.~\ref{fig:StochasticSim}b) and the integrated power at the resonator $P_4$, the open waveguide $P_2$ and the transport efficiency $\eta$ as a function of Lorentzian noise power (Fig.~\ref{fig:StochasticSim}c) are all in good qualitative agreement with the experimental results (Figs.~3c, 4b and \ref{fig:LorCentFreqSweep}).


\begin{figure*}
\includegraphics[trim={0cm 0cm 0cm 0cm},clip,width=1.01\textwidth]{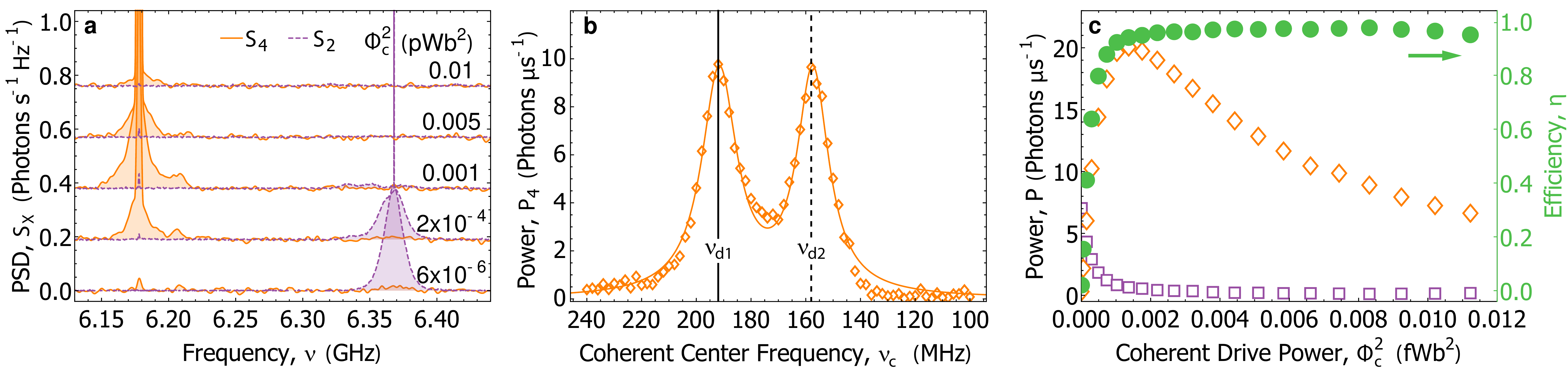} 
\caption{{\bf Coherent environmental modulation of $\Qtwo$ transition frequency.}
\A~Measured power spectral densities of radiation extracted from the resonator $S_4(\omega)$ (solid orange lines) and re-emitted into the transmission line $S_2(\omega)$ (dashed purple lines) for coherent modulation of $\Qtwo$ transition frequency with center frequency at $\nu_\mathrm{c} = 190$~MHz and indicated powers $\Phi_\mathrm{c}^2$. \B~Integrated extracted power $P_4$ as a function of coherent modulation frequency $\nu_\mathrm{c}$ at fixed modulation tone power $\Phi_\mathrm{c}^2 = 54.6$~aWb$^2$. \C~Integrated powers $P_2$ and $P_4$ and the transfer efficiency $\eta$ as a function of coherent modulation power $\Phi_\mathrm{c}^2$.}
\label{fig:CoherentFLInt}
\end{figure*}

\section{Effective Qubit-Environment coupling}
\label{sec:EffectiveQubitPhononCoupling}
Engineered noise with Lorentzain PSD emulates coupling of $\Qtwo$ to a classical phononic mode at center frequency $\omega_\mathrm{L}$ and spectral width $\Delta \omega_\mathrm{L}$. We can estimate the effective qubit-environment coupling by decomposing the applied flux into a large static and a small fluctuating component $\left[\Phi(t) = \Phi_0 + \Delta\Phi(t)\right]$. Using Eq.~(\ref{eq:QubitTune}) the qubit transition energy can be decomposed as 
\begin{eqnarray}
H_\mathrm{q}/\hbar = \omega(t)\sigma_2^z =\omega_0\sigma_2^z + \sigma_2^z\left.\frac{d\omega}{d\Phi}\right\vert_{\Phi_0}\cdot\Delta\Phi(t).
\end{eqnarray}
By assuming that a phononic environmental mode carries at most a single excitation its harmonic spectrum can be effectively substituted by that of a two-level system 
\begin{eqnarray}
H_\mathrm{q}/\hbar \approx \omega_0\sigma_2^z + K \sigma_2^z \left[\sigma_\mathrm{ph}^x(t) + \sigma_\mathrm{ph}^{x\dagger}(t)\right],
\end{eqnarray}
where $K={d\omega}/{d\Phi}\cdot\Delta\Phi_0$ is an effective qubit-environment coupling constant and $(\sigma_\mathrm{ph}^x)$ is dimensionless Pauli operator with unit magnitude.

For experiments with Lorentzian noise we estimate the effective qubit-environment coupling constant ($K$) as a root-mean-square of $\Qtwo$ transition frequency fluctuation induced by applied structured noise.

\section{Modulating the Transition Frequency of $\Qtwo$ with a Coherent Tone}
\label{sec:CoherentMod}

In order to elucidate the mechanism of energy transport for Lorentzian noise applied to $\Qtwo$ we perform an additional measurement in which the $\Qtwo$ transition frequency is coherently modulated via the flux line while it is simultaneously coherently driven via the waveguide. This corresponds to the limiting case of very narrow noise power spectral density. We initially adjust the frequency of the coherent tone $\nu_\mathrm{c}$ to be equal to the $\bright$, $\done$ frequency difference $\Delta_\mathrm{b,d1}$ as in the Lorentzian noise case.

Measurements of the power spectral density at the resonator $S_\mathrm{4}(\omega)$ show, similarly to the Lorentzian noise case (see Fig.~\ref{fig-3}c), a pronounced resonance at $\done$ frequency composed of a broad part with linewidth of approx.~20~MHz and a strong narrow peak with the linewidth of approx.~500~kHz (see Fig.~\ref{fig:CoherentFLInt}a). The narrow peak is comparable in width with the environmental bright $\bright$ state pure dephasing rate $\gamma^\mathrm{b}_\phi/2\pi = 380$~kHz which indicates that it probably originates from the broadened and frequency shifted coherent microwave drive tone applied to the waveguide. 
When sweeping $\nu_\mathrm{c}$ between 100 and 250~MHz and keeping the power of the modulation tone constant at $\Phi_\mathrm{c}^2 = 54.6 \ \mathrm{aWb}^2$, $P_\mathrm{4}$ shows similar dependence as in the white noise case where the extracted power is maximized for $\nu_\mathrm{c} = \Delta_\mathrm{b,d1}$ or $\Delta_\mathrm{b,d1}$ (Fig.~\ref{fig:CoherentFLInt}b). The linewidth of the resonances $\Delta\nu_\mathrm{d1} = 14.1$~MHz and $\Delta\nu_\mathrm{d2} = 15.7$~MHz correspond to the $\done$ and $\dtwo$ state spectral widths. In the case of Lorentzian noise these were additionally broadened by the Lorentzian PSD width of $\Delta\nu_\mathrm{L} = 10$~MHz, which resulted in $\Delta\nu_\mathrm{d1,L} = 30.0$~MHz and $\Delta\nu_\mathrm{d2,L} = 21.5$~MHz (see fit in Fig.~\ref{fig:LorCentFreqSweep}).

For the low frequency coherent modulation the integrated power extracted from the resonator $P_\mathrm{4}$ for $\nu_\mathrm{c} = \Delta_\mathrm{b,d1}$ is almost twice as large as in the Lorentzian noise case (see Fig.~\ref{fig:CoherentFLInt}c and Fig.~\ref{fig-4}b), with approximately half of the power originating from the narrow peak at $\done$. The enhanced value of $P_4$ is in agreement with the model proposed in the main text. The integrated power of the radiation re-emitted into the waveguide $P_\mathrm{2}$ is significantly smaller ($0.7$~Photons/$\mu$s) when $P_4$ reaches its maximum, as compared to white noise (Fig.~\ref{fig-4}a) or Lorentzian noise (Fig.~\ref{fig-4}b) case. As a result the internal transfer efficiency, as defined in the main text, reaches maximum values above 95\%. Although not relevant for light-harvesting processes the depletion of the bright $\bright$ state population is a result of coherent population trapping in the $\done$ state and electromagnetic induced transparency (EIT) of the bright $\bright$ state. In our experiment EIT originates from the destructive interference between coherent excitation of the bright state $\bright$ and strong coherent exchange between $\done$ and $\bright$ state due to low frequency coherent modulation of $\Qtwo$ transition frequency \cite{Abdumalikov2010,Novikov2016}.

\section{Excitation with Incoherent Microwave Radiation}
\label{sec:IncoherentExcitation}

\subsection{Engineering Incoherent Radiation}
\label{sec:IncoherentExcitationA}
\begin{figure}[b]
\includegraphics[trim={0cm 0cm 0cm 0cm},clip,width=0.42\textwidth]{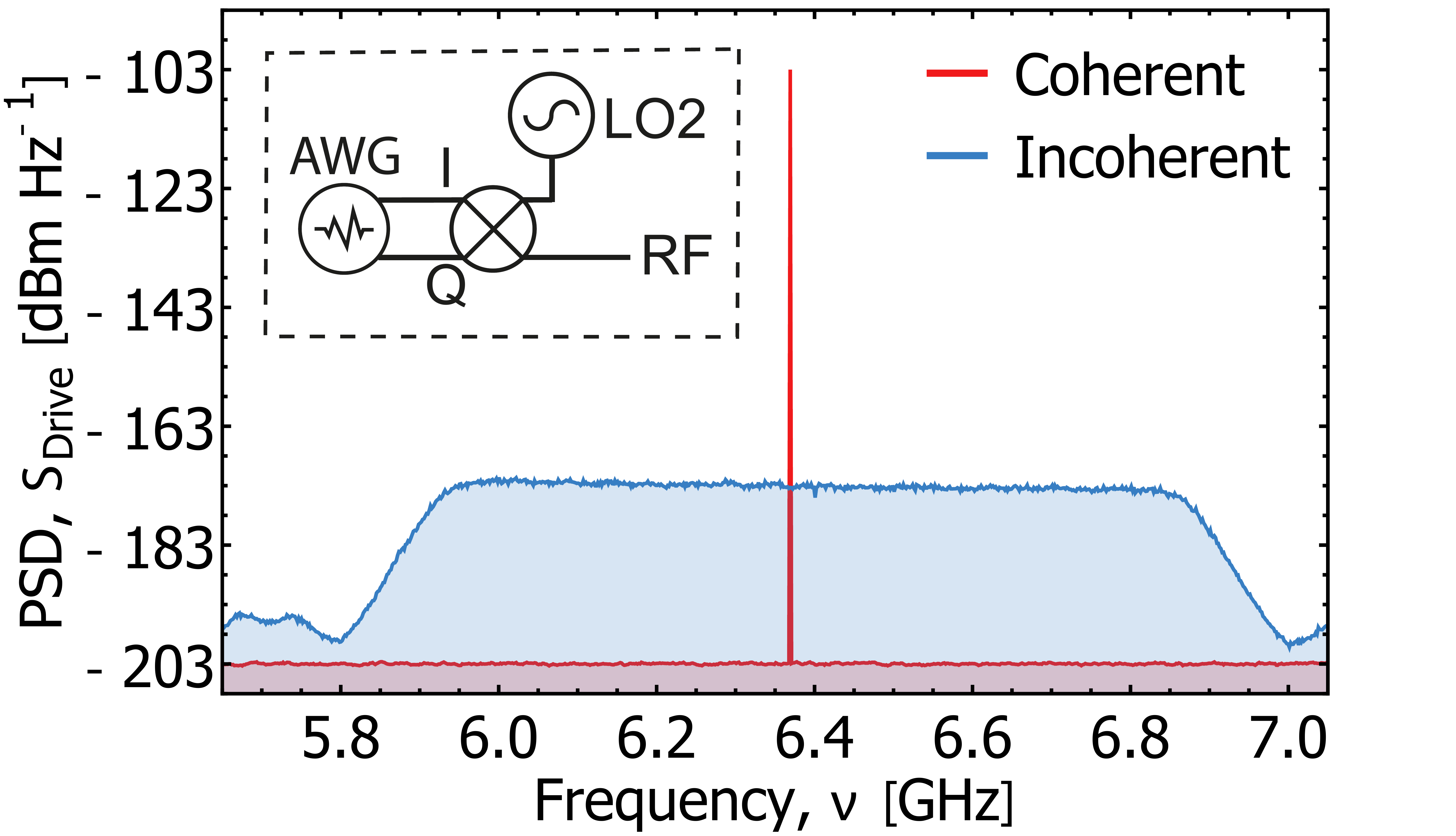} 
\caption{{\bf Power spectral density of applied microwave signals.}
Comparison between incoherent (blue) and coherent microwave signal (red), measured with a spectrum analyzer. The inset shows a diagram of the up-conversion process.}
\label{fig:Incoherent}
\end{figure}

We engineer a broadband incoherent microwave signal by up-converting white noise (see App.~\ref{sec:Noise}). In the up-conversion process the high-frequency LO2 tone is multiplied by a low-frequency signal generated with an AWG (see inset to Fig.~\ref{fig:Incoherent}).
In our experiment, low-frequency white noise with flat spectral density up to $\omega_\mathrm{c}/2\pi = 450$~MHz and exponential cutoff with characteristic width $\Delta\omega/2\pi = 5.44$~MHz is up-converted using a coherent tone at $\omega_\mathrm{LO2}/2\pi = 6.371$~GHz. The resulting high-frequency incoherent signal has a power spectral density with a constant amplitude that spans over a 900~MHz wide band centered at $\omega_\mathrm{B}/2\pi = $~6.371~GHz as shown in Fig.~\ref{fig:Incoherent}. The attenuated incoherent signal is  applied to the sample at port 1, similar to the RF line in Fig.~\ref{fig:ExpSetupDiagram}.

\begin{figure*}[t]
\includegraphics[trim={0cm 0cm 0cm 0cm},clip,width=1.0\textwidth]{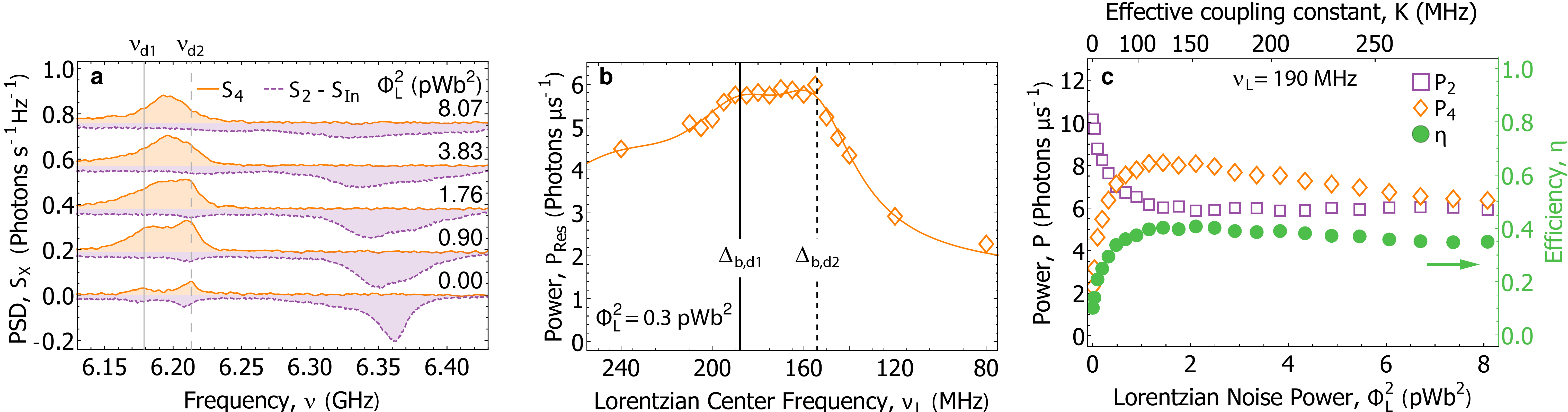} 
\caption{{\bf Incoherently excited system subject to Lorentzian noise.}
\A~Power spectral density detected at the resonator $S_4(\omega)$ (solid orange line) and at the waveguide $S_2(\omega)$ subtracted by separately measured incoherent radiation PSD (dashed purple line) as a function of Lorentzian noise power. \B~Integrated extracted power $P_4$ as a function of the center frequency  $\nu_\mathrm{c}$ of the Lorentzian noise at fixed noise power of $\Phi_\mathrm{L}^2 = 0.3$~pWb$^2$. Solid orange line is a guide to the eye. \C~Integrated powers $P_2$ and $P_4$ and the transfer efficiency $\eta$ as a function of Lorentzian noise power $\Phi_\mathrm{L}^2$ or equivalently the effective qubit-environment coupling constant $K$. The Lorentzian spectrum is centered at $\nu_\mathrm{c} = 190$~MHz for \A~and \B. }
\label{fig:IncoherentResults}
\end{figure*}

\subsection{Incoherently Excited System Subject to Lorentzian Noise}
\label{sec:IncoherentExcitationLorentzian}

Applying Lorentzian noise to the incoherently excited system with central frequency $\nu_\mathrm{L} = 190$~MHz set at the $\bright$-$\done$ frequency difference we observe no enhancement of the extracted power $S_4(\omega)$ at $\nu_\mathrm{d1}$ relative to $\nu_\mathrm{d2}$ (Fig.~\ref{fig:IncoherentResults}a) as in the case of coherent excitation. 
Contrary to the coherent excitation where a multi-photon process is fixed in frequency by a coherent tone, for the incoherent excitation a multi-photon process occurs over a larger frequency range and therefore does not produce a pronounced peak at the $\bright$-$\done$ frequency difference.

The Lorentzian noise increases the transferred power $P_4$ when resonant with the $\dark$ and $\done$ or $\dtwo$ frequency difference (Fig.~\ref{fig:IncoherentResults}b) similar to the coherent excitation case. The transfer efficiency $\eta$ (Fig.~\ref{fig:IncoherentResults}c) shows a non-monotonic behaviour as a function of applied Lorentzian noise power $\Phi_\mathrm{L}^2$ or effective coupling constant $K$ with the maximum at $K/2\pi \approx 130$~MHz, similar to the coherent excitation case.
However, the maximal efficiency $\eta_\mathrm{L,inc.}^\mathrm{max} = 41\%$ is considerably lower compared to the coherent excitation and Lorentzian noise. 
The reduced efficiency can be attributed to the absence of a resonant multi-photon process, which increases the efficiency  of the coherently excited system.
On the other hand, the maximal efficiency obtained with Lorentzian noise ($\eta_\mathrm{L,inc.}^\mathrm{max}$) is larger than the maximal efficiency obtained with white noise ($\eta_\mathrm{W,inc.}^\mathrm{max}$). 
This is in agreement with observations for the coherently excited qubit system. We conclude that the narrow Lorentzian noise spectrum enhances the excitation transport when resonant with the appropriate energy level mismatch for both coherently and incoherently excited qubit systems.

\newpage



\begingroup
\def\refname{References}
\def\bibname{References}

\endgroup

\end{document}